\documentstyle[12pt]{report}
\def\thebibliography#1{\leftline{\it References}\list
  {[\arabic{enumi}]}{\settowidth\labelwidth{[#1]}\leftmargin\labelwidth
    \advance\leftmargin\labelsep
    \usecounter{enumi}}
    \def\newblock{\hskip .11em plus .33em minus .07em}
    \sloppy\clubpenalty4000\widowpenalty4000}

\newcommand{\nonumsection}[1] {\vspace{12pt}\noindent{\bf #1}
        \par\vspace{5pt}}

\newcommand{\be}{\begin{eqnarray}}
\newcommand{\ee}{\end{eqnarray}}
\newcommand{\dslash}{\partial \hskip -0.5em /}
\newcommand{\Dslash}{D \hskip -0.7em /}
\newcommand{\Vslash}{V \hskip -0.7em /}

\newcommand{\Aslash}{A \hskip -0.7em /}
\newcommand{\tr}{{\rm tr}}
\newcommand{\Tr}{{\rm Tr}}

\newcommand{\A}{{\cal A}}

\newcommand{\textlineskip}{\baselineskip=14pt}
\newcommand{\smalllineskip}{\baselineskip=12pt}

\begin{document}

\rightline{UNITU-THEP-13/1994}
\rightline{July 1994}
\rightline{hep-ph/9407304}
%\vskip .5truecm
%\rightline{\it PACS: 11.10.LM, 14.20.-c}
\vskip 0.8truecm
\centerline{\Large\bf On the analytic properties of chiral solitons}
\vskip 0.4truecm
\centerline{\Large\bf in the presence of the $\omega$--meson}
\baselineskip=20 true pt
\vskip 0.8cm
\centerline{H.\ Weigel$^{\dag}$, U.\ Z\"uckert, R.\ Alkofer, and
H.\ Reinhardt}
\vskip .3cm
\centerline{Institute for Theoretical Physics}
\centerline{T\"ubingen University}
\centerline{Auf der Morgenstelle 14}
\centerline{D-72076 T\"ubingen, Germany}
\vskip 0.6cm
\baselineskip=16pt
\centerline{\bf ABSTRACT}
\vskip .2cm
A thorough study is performed of the analytical properties of the
fermion determinant for the case that the time components of
(axial) vector fields do not vanish. For this purpose the
non--Hermitian Euclidean Dirac Hamiltonian is generalized to the
whole complex plane. The Laurent series are proven to reduce to
Taylor series for the corresponding eigenvalues and --functions as
long as field configurations are assumed for which level crossings do
not occur. The condition that no level crossings appears determines
the radius convergence. However, the need for regularization prohibits
the derivation of an analytic energy functional because real and
imaginary parts of the eigenvalues are treated differently. Consistency
conditions for a Minkowski energy functional are extracted from global
gauge invariance and the current field identity for the baryon current.
Various treatments of the Nambu--Jona--Lasinio soliton are examined
with respect to these conditions.

Motivated by the studies of the Laurent series for the energy functional
the Euclidean action is expanded in terms of the $\omega$--field.
It is argued that for this expansion the proper--time regularization
scheme has to be imposed on the operator level rather than on an
expression in terms of the one--particle eigenenergies. The latter
treatment is plagued by the inexact assumption that the Euclidean
Dirac Hamiltonian and its Hermitian conjugate can be diagonalized
simultaneously. It is then evident that approaches relying on counting
powers of the $\omega$--field in the one--particle eigenenergies are
inappropriate.

Using the expansion of the action up to second order and
employing a parametrical description of the soliton profiles the
repulsive character of the $\omega$ meson is confirmed. In the
presence of the $\omega$ meson the soliton mass is enhanced by
a few hundred MeV.

\vfill
\noindent
$^{\dag}$
{\footnotesize{Supported by a Habilitanden--scholarship of the
Deutsche Forschungsgemeinschaft (DFG).}}
\eject

\normalsize\textlineskip
\stepcounter{chapter}
\leftline{\large\it 1. Introduction and Motivation}
\bigskip

In the past few years the picture of baryons as being soliton
solutions in effective meson theories has proven to be quite
successful \cite{ho93}. As an example we refer to the appealing
explanation of the almost vanishing matrix element of the axial
singlet current for the proton, {\it i.e.} the so--called
``spin--puzzle" \cite{br88}.

The investigation of the baryon structure in soliton models
was started off by the consideration of the large $N_C$ (number of
color degrees of freedom) limit \cite{tho74} of Quantum Chromo Dynamics
(QCD), the theory generally accepted to describe processes being
subject to the strong interacting. In this limit QCD was shown to
be equivalent to an effective theory of weakly (${\cal O}(1/N_C)$)
interacting mesons. Witten furthermore conjectured that baryons emerge
as (topological) solitons of this meson theory \cite{wi79}. At small
energies (or large distances) this effective theory should only
contain the low--lying meson states, {\it e.g.} the pseudoscalar
would--be Goldstone boson of chiral symmetry ($\pi,K,\eta$). A further
requirement when modeling QCD was to maintain its symmetries,
especially the chiral symmetry and its spontaneous breaking.
These ideas were brought to practice by the rediscovery of the Skyrme
model \cite{sk61} by Adkins, Nappi and Witten \cite{ad83}. The Skyrme
model represents an effective theory of the pseudoscalar pions (and
kaons in the three flavor generalization) and admits topological
soliton solutions. In the spirit of Witten's conjecture the baryon
number current, $B_\mu$, was identified with the topological current
\cite{wi83}.

Soon after this, vector mesons were added to the scenery in a way
consistent with chiral symmetry \cite{ka85,sch86,ba87}. This has turned out
to be especially pleasing since it brought together the fruitful concept
of vector meson dominance (VMD) and the soliton picture of baryons
\cite{me87}. In particular the isoscalar--vector $\omega$--meson was
shown to play a crucial role. It is directly connected to the baryon
current via the stationary condition which generically reads
\be
\left(D^{\mu\nu}+m_\omega^2g^{\mu\nu}\right)\omega_\nu=gB^\mu.
\label{e1omega}
\ee
Here $D^{\mu\nu}$ denotes a kinetic operator ({\it e.g.}
$\partial^\mu\partial^\nu-g^{\mu\nu}\partial_\rho\partial^\rho$), which
depends on the specific form of the model and $g$ is a coupling
constant. For the time component, $\mu=0$, eqn (\ref{e1omega}) is rather
a constraint than an equation of motion. VMD then implies to identify
the isoscalar part of the electromagnetic current as \cite{sa69}
\be
j^\mu_{I=0}=\frac{e m_\omega^2}{g}\omega^\mu
\label{i0current}
\ee
with $e$ being the electric charge\footnote{As the integral over
$D^{\mu\nu}\omega_\nu$ vanishes $j^\mu_{I=0}$ is properly
normalized.}. Transforming (\ref{e1omega}) to momentum
space it is straightforward to verify that the isoscalar
radius of the nucleon is given by \cite{me87}
\be
r^2_{I=0}=r^2_B+\frac{6}{m_\omega^2}
\label{i0radius}
\ee
where $r^2_B=\int d^3r r^2 B_0$ denotes the intrinsic baryonic radius.
The additional contribution $6/m_\omega^2$, which is completely due
to VMD, is crucial in order to reproduce the experimental value
for $r^2_{I=0}\approx(0.86{\rm fm})^2$.

More recently efforts have been made to derive the effective meson
theory from QCD motivated models of the quark flavor dynamics. In
this context the model of Nambu and Jona--Lasinio (NJL) \cite{na61} has
acquired special attention \cite{ha94}. The associated Lagrangian
\be
{\cal L}_{\rm NJL} = \bar q (i\dslash -\hat m^0 ) q
& + & 2G_1 \sum _{i=0}^{N_f^2-1}
\left( (\bar q \frac {\lambda ^i}{2} q )^2
      +(\bar q \frac {\lambda ^i}{2} i\gamma _5 q )^2 \right)
\nonumber \\*
 & - & 2G_2 \sum _{i=0}^{N_f^2-1}
\left( (\bar q \frac {\lambda ^i}{2} \gamma _\mu q )^2
      +(\bar q \frac {\lambda ^i}{2} \gamma _5 \gamma _\mu q )^2 \right)
\label{njl}
\ee
is chirally symmetric. Here $q$ denotes the quark spinors, $\hat m^0$ the
current quark mass matrix and $N_f$ the number of flavor degrees of
freedom. The latter will be taken to be two from now on. As the
coupling constants $G_i$ carry dimension $({\rm mass})^{-2}$ the model
is not renormalizable. This requires regularization and introduces
an additional parameter, the cut--off $\Lambda$. For a sufficiently
large coupling $G_1$ the model indeed exhibits spontaneous
breaking of chiral symmetry \cite{na61}. This feature is reflected
by a non--vanishing quark condensate $\langle\bar q q\rangle$.

The Lagrangian (\ref{njl}) can be rewritten in terms of composite
meson fields by functional integral techniques \cite{eb86}. This
procedure is referred to as bosonization. In order to make these
functional integrals well behaved it is mandatory to impose
Feynman boundary conditions or, equivalently, to perform these
integrals in Euclidean space. In Euclidean space the time--coordinate
is analytically continued to $x_0=-ix_4=-i\tau$. It also implies
to continue the time components of the vector
($V_\mu\sim\bar q\gamma_\mu q$) and axial vector
($A_\mu\sim\bar q\gamma_\mu\gamma_5 q$) fields
\be
V_0\rightarrow iV_4,\quad A_0\rightarrow iA_4
\label{veccont}
\ee
and consider $\tau,V_4$ and $A_4$ as Hermitian quantities. In Euclidean
space the bosonized action finally is given as the sum of a purely
mesonic term
\be
\A_{\rm m} = \int d^4x \left( -\frac 1 {4G_1}\tr
(\Sigma^{\dag}\Sigma-\hat m^0(\Sigma+\Sigma^{\dag})+(\hat m^0)^2)
+\frac{1}{4G_2}\tr(V_\mu V^\mu+A_\mu A^\mu) \right)
\label{mesact}
\ee
and a fermion determinant
\be
\A_{\rm F} = \Tr\ \log (i\Dslash_E )
 = \Tr\ \log \big(i\dslash +\Vslash +\gamma_5\Aslash
- (P_R\Sigma+P_L\Sigma^{\dag}) \big)
\label{feract}
\ee
{\it i.e.} $\A_{\rm NJL}=\A_{\rm F}+\A_{\rm m}$.
Here $P_{R,L} = (1\pm \gamma _5)/2$ are the projectors on right-- and
left--handed quark fields, respectively. The complex field $\Sigma$
describes the scalar and pseudoscalar meson fields,
$S_{ij}= S^a\tau^a_{ij}/2$ and $P_{ij}= P^a\tau^a_{ij} /2$:
$\Sigma=S + i P$. There is no problem with treating $\A_{\rm m}$ as
it obviously is analytical in the meson fields. This, however, is not
the case for $\A_{\rm F}$. As a matter of fact $\A_{\rm F}$ is (in
Euclidean space) not a real quantity and thus may be decomposed into
real ($\A_R$) and imaginary ($\A_I$) parts
\be
\A_R=\frac{1}{2}\Tr\ {\rm log}\left
(\Dslash_E\Dslash_E\hspace{0.04cm}^{\dag}\right),
\qquad
\A_I=\frac{1}{2}\Tr\ {\rm log}\left(\Dslash_E
(\Dslash_E\hspace{0.04cm}^{\dag})^{-1}\right).
\label{arai}
\ee

The above mentioned ultraviolet divergencies are completely contained
in $\A_R$.\footnote{In order to obtain a finite value for $\A_I$ a
suitable definition of the functional trace is mandatory. See section
3 for details.} Again the treatment in Euclidean space proves to be
pertinent since the operator
$\Dslash_E\Dslash_E\hspace{0.04cm}^{\dag}$ is
positive definite. Thus the proper--time regularization \cite{sch51}
prescription may be applied to $\A_R$ resulting in the replacement
\be
\A_R\longrightarrow-\frac{1}{2}{\rm Tr}
\int_{1/\Lambda^2}^\infty\frac{ds}{s}\ {\rm exp}
\left(-s\Dslash_E\Dslash_E\hspace{0.04cm}^{\dag}\right).
\label{arreg}
\ee

Expanding the fermion determinant in terms of the meson fields and
their derivatives leads to an effective meson theory \cite{eb86}. As
this object is, by construction, a polynomial in the meson fields the
analytic continuation back to Minkowski space is well defined. The
resulting meson theory reasonably well describes the physics of the
pseudoscalar and vector mesons. Furthermore this theory has many
features in common with Skyrme type models which are supplemented by
vector mesons.

This effective theory also allows one to express physical quantities
like the pion decay constant $f_\pi$ in terms of the cut--off $\Lambda$
and the constituent quark mass\footnote{The difference
$m-m_0=-2G_1\langle\bar q q\rangle$ reflects the spontaneous
breaking of chiral symmetry.} $m=\langle\Sigma\rangle$.
In case the axial meson degrees of freedom are ignored this
relation reads in the gradient expansion \cite{eb86}
\be
f_\pi^2=\frac{N_C m^2}{4\pi^2}
\Gamma\left(0,\left(\frac{m}{\Lambda}\right)^2\right).
\label{fpi}
\ee
In practice the physical value $f_\pi=93$MeV will be employed
to determine $\Lambda$ for a given constituent quark mass $m$.

Later on it was also demonstrated that the bosonized version of the
NJL model contained static soliton solutions which extremize
the Minkowski energy functional \cite{re88}. For the investigation
of static field configurations the introduction of a Dirac Hamiltonian
in Euclidean space, $h_E$, via
\be
i\beta\Dslash_E=-\partial_\tau-h_E
\label{edirac}
\ee
is useful because $[\partial_\tau,h_E]=0$. The Euclidean energy
functional is then expressed in terms of the eigenvalues of $h_E$.
First calculations were constrained to the pseudoscalar mesons.
Subsequently the isovector--vector \cite{al91} and axial vector
mesons \cite{al92} were added. The associated field configurations
have vanishing temporal components $V_4=0$ and $A_4=0$ rendering
$h_E$ an Hermitian operator. Thus the extraction of a Minkowski energy
functional from the one in Euclidean space is straightforward. In
this context it should be remarked that especially the inclusion of
the axial vector meson provides further support for Skyrme type
models because it turns out that then the valence quark orbit joins
the Dirac sea and the baryon number is completely carried by the
distorted Dirac vacuum \cite{al92}.

The analytic structure of the energy functional changes
drastically when the isoscalar vector $\omega$--meson is included. In
this case the soliton configuration involves a non--vanishing temporal
component leading to a non--Hermitian operator $h_E$. So far three
distinct approaches to include the $\omega$--meson into the static
energy functional of the NJL model have been discussed in the literature
\cite{wa92}--\cite{zu94}. It is the main purpose of the present paper to
illuminate the role of the $\omega$--meson in the NJL soliton model
and to study the behavior of the corresponding fermion determinant
under transformations which mediate between Euclidean and Minkowski
space. The above mentioned importance of the $\omega$--meson for the
description of static baryon properties requires a clarification
of the situation.

The remainder of this paper is organized as follows. In section 2 we
investigate the analytical properties of the eigenvalues and --vectors
of $h_E$. In section 3 we briefly review the derivation of the
Euclidean energy functional and show that the analytic continuation
which should transform this object to Minkowski space does not
exist. In section 4 requirements on the Minkowski energy functional
from a phenomenological point of view are derived. Furthermore
several definitions for the Minkowski energy functional are
discussed. Numerical simulations for the eigenvalues of $h_E$ show
that treating the $\omega$--meson up to quadratic order represents a
reasonable approximation. Using this result, it is argued that the
same approximation for the Euclidean energy functional is pertinent.
Then a Minkowski energy functional can actually be derived. The study
of this approximation is presented in section 5. In that section we
also demonstrate that counting powers of the $\omega$--field in the
eigenvalues of $h_E$ leads to incorrect results for the module of the
fermion determinant. As a matter fact, these misinterpretations are
related to the appearance of $\Dslash_E^{\dag}$ in the regularized
version of the fermion determinant. Concluding remarks may be found
in section 6. Technical details on the perturbative expansion used in
section 5 are discussed in an appendix.

\bigskip
\stepcounter{chapter}
\leftline{\large\it 2. Analytic Properties of the Dirac Hamiltonian}
\bigskip

In the introductory section we have argued that it is mandatory
to analytically continue forth and back from Minkowski to Euclidean
spaces in order to obtain a meaningfully regularized energy functional.
In the present section we will discuss the consequences of this
continuation on the eigenvalues and --vectors of the static
Dirac Hamiltonian $h_E$ which can be extracted from eqn (\ref{edirac}).

As inferred from eqn (\ref{veccont}) the time components of the
(axial) vector meson play the key role for the study of the analytic
properties. The {\it ans\"atze} for the static meson field
configurations are characterized by the fact that they commute with
the grand spin operator $\mbox{\boldmath $G$}$. This operator denotes
the sum of total angular momentum (orbital plus spin) and isospin,
{\it i.e.} $\mbox{\boldmath $G$}=
\mbox{\boldmath $l$}+\mbox{\boldmath $\sigma$}/2
+\mbox{\boldmath $\tau$}/2$. Demanding in addition the proper behavior
under parity the time component of the axial vector has to
be zero while that of the vector field reduces to purely radial
function $\omega(r)$. In Euclidean space the Dirac Hamiltonian may
then be decomposed as
\be
h_E=h_\Theta+i\omega(r).
\label{decomph}
\ee
Here $h_\Theta$ denotes the Hermitian part of $h_E$, {\it i.e.}
$h_\Theta^{\dag}=h_\Theta$. Let us, for the time being, consider the
simplest case when only pseudoscalar fields are present. The
corresponding field configuration is assumed to be of hedgehog type
\be
\Sigma=m\ {\rm exp}\left(i\hat{\mbox{\boldmath $r$}} \cdot
\mbox{\boldmath $\tau$}\Theta(r)\right).
\ee
Hence
\be
h_\Theta=\mbox{\boldmath $\alpha$}\cdot\mbox{\boldmath $p$}
+m\beta\left({\rm cos}\Theta+i\gamma_5\hat{\mbox{\boldmath $r$}}
\cdot\mbox{\boldmath $\tau$} {\rm sin}\Theta\right).
\label{htheta}
\ee
Later on we will also consider the case when $\rho$ and $a_1$ fields are
present as well. However, for the general discussion of the analytical
properties the restriction to $\Theta$ and $\omega$ proves to be
illuminating. For this discussion we generalize the Dirac Hamiltonian
to an operator which depends on the complex variable $z$
\be
h(z)=h_\Theta+z\omega.
\label{hz}
\ee
The Euclidean space Hamiltonian, $h_E$ then corresponds to $z=i$
while the one in Minkowski space is associated with $z=1$. We will
refer to these values of $z$ as Euclidean and Minkowski points,
respectively. Except for the Minkowski point as well as $z=-1$, $h(z)$
is non--Hermitian. In general we therefore have to distinguish between
left ($\Psi_\nu(z)$) and right ($\tilde{\Psi}_\nu(z)$) eigenstates
\be
h(z)|\Psi_\nu(z)\rangle & = & \epsilon_\nu(z) |\Psi_\nu(z)\rangle
\label{eigen} \\*
\langle\tilde{\Psi}_\nu(z)|h =
\epsilon_\nu(z)\langle\tilde{\Psi}_\nu(z)|
\quad & \Longrightarrow & \quad
h^{\dag}(z) |\tilde{\Psi}_\nu(z)\rangle = \epsilon_\nu(z)^*
|\tilde{\Psi}_\nu(z)\rangle.
\nonumber
\ee
Of course, the eigenvalues, $\epsilon_\nu(z)$, and --vectors,
$|\Psi_\nu(z)\rangle$, parametrically depend on $z$. Adopting the
set of free spherical grand spinors as basis \cite{ka84} the matrix
elements of $h_\Theta$ and $\omega$ turn out to be symmetric. Since
$h^{\dag}(z)=h(z^*)$ the eigenvalues and their complex conjugate are
related via
\be
\epsilon_\nu(z)^*=\epsilon_\nu(z^*).
\label{estar}
\ee
Furthermore a phase convention may by chosen such that the
wave--functions are related by
\be
|\tilde\Psi(z)\rangle=|\Psi(z)^*\rangle=|\Psi(z^*)\rangle.
\label{psistar}
\ee
For later discussions it will be useful to define real and imaginary
parts of the eigenvalues
\be
\epsilon^R_\nu(z,z^*)
=\frac{1}{2}\left(\epsilon_\nu(z)+\epsilon_\nu(z^*)\right)
\qquad
\epsilon^I_\nu(z,z^*)
=-\frac{i}{2}\left(\epsilon_\nu(z)-\epsilon_\nu(z^*)\right)
\label{erei}
\ee

It is the main issue of this section to explore the $z$--dependence
of the eigenvalues and --vectors once profile functions
$\Theta(r)$ and $\omega(r)$ are made available. Of course, these
test profiles can be chosen arbitrarily, however, they should
(at least) be physically motivated. For convenience we will employ
the profile functions displayed in figure 5.2 and
constrain ourselves to the case $m=400$MeV. The reader may
consult section 5 on the origin of these profiles. Here it suffices
to remark that the $\omega$ field satisfies an equation analogous
to (\ref{e1omega}) and is thus properly normalized.
In order to study the behavior of the eigenvalues (and --vectors)
in the complex plane defined by $z$ we parametrize
\be
z=\delta{\rm exp}\left(i\varphi\right)\quad
{\rm with}\quad 0\le\varphi\le2\pi.
\label{zpar}
\ee
Obviously $\delta=1$ and $0\le\varphi\le\pi/2$ parametrizes the
continuation from Minkowski to Euclidean space and vice versa.
However, we will consider $\delta$ as a parameter in order to
determine the radius of convergence in the complex $z$--plane.
{\it A priori} it is not obvious that the energy eigenvalues exhibit
any kind of analytic structure  as these are roots of the
characteristic polynomial (See {\it e.g.} appendix C of
ref.\cite{zu94}.). This polynomial is in principle of infinite degree
and only the restriction to a finite basis renders this degree finite
as well. In figure 2.1 the dependence on $\delta$ for two different
energy eigenvalues is shown. On the left the energy of the valence quark
orbit is displayed. This orbit is, by definition, the one with the
smallest absolute value of the real part of the energy eigenvalue.
{\it I.e.} the valence quark state is associated with the smallest
$|\epsilon_\nu^R|$ (see also section 3). The valence quark state is
distinct since in the presence of static meson fields it is the
one which deviates most strongly from a free quark state.
Obviously the valence quark energy represents a smooth function
of $\varphi$ for $\delta\le1$ while for $\delta=2$ the derivative
$\partial\epsilon_{\rm val}/\partial\varphi$ becomes singular.
At the ``edges" $\varphi$ assumes the values $\pm0.32\pi$. These
singularities are, however, not the consequence of
a non--analytic behavior of the roots of the characteristic polynomial
but rather a level crossing appears. For $|\varphi|\le0.32\pi$ a
state with negligible $\varphi$--dependence carries the smallest
$|\epsilon_\nu^R|$. For $|\varphi|>0.32\pi$ the role of the valence
quark is then taken over by an orbit which strongly depends on
$\varphi$. Numerically we have confirmed that for the test profiles
under consideration such level crossings in the valence quark
channel are avoided as long as $\delta<1.3$. This radius of
convergence suggests that for the relevant parameter space
($\delta=1$) the eigenvalues of the Dirac Hamiltonian are analytic
functions of $z$. This result was already obtained
previously \cite{sch93tr}. This is furthermore supported by
considering other levels than the valence quark state. These actually
possess a less pronounced dependence on $\omega$ and have thus a larger
radius of convergence. As an example the energy of the state with the
smallest $|\epsilon^R|$ in the channel with grand spin and
parity $G^\pi=2^+$ is displayed in figure 2.1. For this energy the
radius of convergence is obviously larger than two.

There is actually a much more elegant technique to investigate the
analytical structure of the eigenvalues and especially to extract
the radius of convergence. As these eigenvalues are supposedly
analytic in $z$ the corresponding Laurent expansion
\be
\epsilon_\nu(z)=\sum_{n=-\infty}^\infty
c_{\nu,n}(z_0)(z-z_0)^n
\label{lauser}
\ee
should exist. Here $z_0$ serves as the center of the expansion and may
refer to any point in the complex plane. The Laurent coefficients
$c_{\nu,n}(z_0)$ are defined by the Cauchy integrals
\be
c_{\nu,n}(z_0)=\frac{1}{2\pi i}
\oint_{\partial A}\frac{\epsilon_\nu(\zeta)d\zeta}
{\left(\zeta-z_0\right)^{n+1}}
\label{cauchy}
\ee
with $z_0$ being in the interior of the integration contour
$\partial A$, {\it i.e.} $z_0\in A$. Employing the parametrization
$\zeta=\delta e^{i\varphi}+z_0$ these Cauchy integrals become
ordinary integrals
\be
c_{\nu,n}(z_0)=\frac{1}{2\pi\delta^n}
\int_0^{2\pi}\epsilon_\nu\left(\delta e^{i\varphi}+z_0\right)
{\rm exp}\left(-in\varphi\right)d\varphi.
\label{cpara}
\ee
The computation of the integrals (\ref{cpara}) obviously requires
the eigenvalues of
\be
h_\Theta+\left(\delta e^{i\varphi}+z_0\right)\omega
\label{hcauchy}
\ee
in the interval $0\le\varphi\le2\pi$. Analyticity is then equivalent
to the fact that the coefficients $c_{\nu,n}(z_0)$ do actually
${\underline {\rm not}}$ depend on $\delta$. On the other hand the
$c_{\nu,n}(z_0)$ are useful tools to determine the radius of convergence
at $z_0$. For this purpose we start off with a small value for $\delta$,
numerically compute the eigenvalues of (\ref{hcauchy}) and subsequently
the integrals (\ref{cpara}). Next the calculations are repeated assuming
a somewhat larger value for $\delta$. As long as $\delta$ is
sufficiently small the $c_{\nu,n}(z_0)$ are indeed independent of
$\delta$. However, as a critical value is exceeded the
$c_{\nu,n}(z_0)$ vary with $\delta$. This critical value has to be
identified as the radius of convergence at the point $z_0$. For
the valence quark orbit the results of this calculation are displayed
in table 2.1. Again we have employed the test profiles which have been
mentioned above. For the interpretation of this table (and similar
once which will follow) it should be stressed that the entry ``0.00"
means that this value is zero up to the given accuracy. On the other
hand the entry ``0" implies that the corresponding coefficient vanishes
identically up to numerical uncertainties which usually are of the order
$10^{-8}$.
\begin{table}
\caption{The coefficients $c_{{\rm val},n}(z_0)$ in the Laurent series
(\protect\ref{lauser}) for the valence quark energy in units of the
constituent quark mass.}
{}~
\newline
\centerline{\tenrm\smalllineskip
\begin{tabular}{|c|c c c c|c c c c|}
\hline
 &\multicolumn{4}{c|}{$z_0=0$} & \multicolumn{4}{c|}{$z_0=i$} \\
$\delta$ & 0.5  & 1.0 & 1.2 & 1.5 & 0.5 & 1.0 & 1.2 & 1.5\\
\hline
$n=-4$  & 0    &  0    & 0    & 0.18 &
0    & 0    & 0.08-0.07$i$ & 0.22+0.05$i$ \\
$n=-3$  & 0    &  0    & 0    & 0.14 &
0    & 0    & 0.06-0.09$i$ & 0.20-0.12$i$ \\
$n=-2$  & 0    &  0    & 0    & 0.08 &
0    & 0    & 0.04-0.10$i$ & 0.12-0.20$i$ \\
$n=-1$  & 0    &  0    & 0    & 0.02 &
0    & 0    & 0.01-0.10$i$ & 0.02-0.20$i$ \\
$n=0$   & 0.38 &  0.38 & 0.38 & 0.36 &
0.38+0.61$i$ & 0.38+0.61$i$ & 0.39+0.52$i$ & 0.36+0.46$i$ \\
$n=1$   & 0.59 &  0.59 & 0.59 & 0.56 &
0.62-0.03$i$ & 0.62-0.03$i$ & 0.60-0.09$i$ & 0.56-0.11$i$ \\
$n=2$   &-0.02 & -0.02 &-0.02 &-0.05 &
-0.01-0.02$i$ &-0.01-0.02$i$ &-0.03-0.07$i$ &-0.06-0.05$i$ \\
$n=3$   &-0.01 & -0.01 &-0.01 &-0.03 &
 0.00 & 0.00 &-0.03-0.03$i$ &-0.03-0.01$i$ \\
$n=4$   & 0.00 &  0.00 & 0.00 &-0.01 &
0.00 & 0.00 &-0.02-0.02$i$ &-0.01 \\
\hline
\end{tabular}}
\end{table}
Evaluating the Laurent coefficients associated with an expansion
around the origin of the complex plane we conclude that the radius
of convergence for the energy of the valence quark energy eigenvalue
lies in between 1.2 and 1.5. In the region of convergence the Laurent
series actually reduces to a Taylor series without any poles. Then
the analytic continuation from Euclidean to Minkowski space is unique
for this eigenvalue. However, as will be discussed in the next section,
the physical relevant situation is somewhat different. The fermion
determinant is defined in Euclidean space only. Thus one has to
perform a Laurent expansion around $z_0=i$. The results for the
coefficients stemming from this calculation are also shown in
table 2.1 and demonstrate that for this expansion the radius of
convergence is less than 1.2. Taking into account that the distance
between the Euclidean and Minkowski points is $\sqrt2$ we see that
it is not possible to carry out the required continuation for the
valence quark level, at least for the test profiles considered here.
Again this non--analyticity originates from a level crossing as
discussed above. Thus the identification of the valence quark level
does not necessarily symbolize an analytic operation. For orbits
other than the valence quark the continuation is less problematic
because their response to the $\omega$--profile is significantly
weaker. As an example we have again considered the lowest state
in the $2^+$ channel. As can be inferred from table 2.2 the
dependence of the corresponding energy eigenvalue on $z$ is
almost negligible.

\begin{table}
\caption{Same as table 2.1 for the state with the smallest (positive)
real part of the energy eigenvalue in the $G^\pi=2^+$ channel. All
those coefficients which are not displayed either vanish ($n<0$) or
are negligibly small ($n\ge2$).}
{}~
\newline
\centerline{\tenrm\smalllineskip
\begin{tabular}{|c|c c c|c c c|}
\hline
 &\multicolumn{3}{c|}{$z_0=0$} & \multicolumn{3}{c|}{$z_0=i$} \\
$\delta$ & 0.5  & 1.0 & 1.5 & 0.5 & 1.0 & 1.5\\
\hline
$n=0$   & 1.1449 & 1.1449 & 1.1449 &
1.1453+0.0021$i$ & 1.1453+0.0021$i$ & 1.1453+0.0021$i$ \\
$n=1$   & 0.0022 & 0.0022 & 0.0022 &
0.0020-0.0005$i$ & 0.0020-0.0005$i$ & 0.0020-0.0005$i$ \\
\hline
\end{tabular}}
\end{table}

There is one more conclusion which can be drawn from the results
listed in tables 2.1 and 2.2. In the region of convergence only
those Laurent coefficients are sizable which correspond to a constant
or linear dependence of the eigenvalues on $z$. In addition a small but
non--vanishing quadratic piece may be present. This observation may be
translated into the statement that the functional dependence of the
energy eigenvalues on $\omega$ is at most quadratic. Of course, this
again depends on the test profiles. However, our computations
indicate that this is a generic feature for $\omega$--meson profiles
which are in some way related to the quark baryon current and hence
are properly normalized.

We have performed similar investigations not only for the
eigenvalues but also for the eigenstates of
$h(z)$ (see eqn (\ref{eigen})). The symmetry of $h_\Theta$ and
$\omega$ in the standard basis straightforwardly relates left and
right eigenstates by complex conjugation. It is thus sufficient to
only consider the behavior of the right eigenstates in the complex $z$
plane. For the valence quark wave--function we display the dependence on
$z=\delta{\rm exp}(i\varphi)$ in figure 2.2. Again we observe that the
dependence is analytical for $\delta\le1.0$ and that for $\delta$ as
large as 2.0 the identification of the valence quark level exhibits
an arbitrariness. Although the somewhat shaky curve for the lower
component seems to suggest some kind of non--analyticity even
for $\delta=1.0$ this is actually not the case as we have verified
by computing the Laurent series in analogy to (\ref{lauser}).

At the end of this section we thus arrive at the statement that
except for the identification of the valence quark state the
diagonalization of $h(z)$ results in analytic expressions for its
eigenvalues and --functions. For the identification of the valence
quark state the radius of convergence is comparable to the distance
between Euclidean and Minkowski points and depends on the center of
the Laurent expansion. The identification will not cause problems
if the self--consistent $\omega$--meson profiles turn out to be less
pronounced than the test profiles considered here. Having computed a
self--consistent solution it will be necessary to verify that
the corresponding profile functions do not cause any level crossing
for the valence quark state and the associated non--analytic behavior
is avoided on the path from Euclidean to Minkowski spaces. Otherwise
the solution has to discarded.

\newpage
\bigskip
\stepcounter{chapter}
\leftline{\large\it 3. Analytic ``Properties" of the Energy Functional}
\bigskip

In the previous section we have examined the analytic properties
of the eigenvalues and --states of the one particle Dirac Hamiltonian
(\ref{hz}) in the complex $z$--plane. Here we will investigate how
(or whether at all) these properties propagate to the energy functional.
For this purpose we will briefly review the derivation of the energy
functional in Euclidean space and also emphasize on an important issue
when evaluating the imaginary part of the action, ${\cal A}_I$, see
eqn (\ref{arai}).

Ignoring for the moment the regularization of the fermion
determinant associated with the Hamiltonian (\ref{hz}) it can
formally be written as \cite{re89,al94}
\be
{\rm Det}\left(\Dslash_E\right)={\cal C}(T){\rm exp}(-TE_0)
\sum_{\{\eta_\nu\}}{\rm exp}
\left(-T N_C\sum_{\nu}\eta_\nu\bar{\epsilon}_\nu\right).
\label{dete}
\ee
Here $T$ denotes the Euclidean time interval under consideration.
${\cal C}(T)$ refers to a constant of proportionality which does not
depend on the eigenvalues $\epsilon_\nu$. Hence its explicit form
is of no further relevance for our present investigations.
The sum goes over all possible configurations $\{\eta_\nu\}$ of
(anti) quark occupation numbers $\eta_\nu=0,1$. The quantities
${\bar\epsilon_\nu}$ are defined in terms of the real and
imaginary parts of the eigenvalues of $\epsilon_\nu$ (see
eqn (\ref{erei}))
\be
{\bar\epsilon_\nu}=|\epsilon_\nu^R|
+i\ {\rm sgn}(\epsilon_\nu^R)\epsilon_\nu^I
={\rm sgn}(\epsilon_\nu^R)\epsilon_\nu.
\label{ebar}
\ee
The limit $T\rightarrow\infty$ projects out the vacuum energy
\be
E_0=-\frac{N_C}{2}\sum_\nu{\bar\epsilon_\nu}
\label{evacnonreg}
\ee
from the exponential. Eqn. (\ref{dete}) provides a natural definition
of the valence quark part of the energy
\be
E_V=E_V^R+iE_V^I=
N_C\sum_\nu\eta_\nu |\epsilon_\nu^R|+
iN_C\sum_\nu\eta_\nu {\rm sgn}(\epsilon_\nu^R)
\epsilon_\nu^I.
\label{eeval}
\ee
It should be noted that the unregularized fermion determinant
(\ref{dete}) actually is a function of the energy eigenvalues
$\epsilon_\nu=\epsilon_\nu^R+i\epsilon_\nu^I$ rather than of their
real and imaginary parts separately. Thus the complete fermion
determinant ({\it i.e.} when the sum over all sets of occupation
numbers is carried out) would represent an analytic function of
$z$ if it were finite and as long as the eigenvalues were analytic
functions of $z$. For practical calculations, however, one considers
only configurations with a definite baryon number
\be
B=\sum_\nu\left(\eta_\nu-\frac{1}{2}\right)
{\rm sgn}(\epsilon_\nu^R).
\label{defb}
\ee
This corresponds to restrict oneself to a special set of
occupation numbers without carrying out the associated sum.
Then an explicit dependence on a subset of the quantities
$\bar\epsilon_\nu$ will induce a non--analyticity if
${\rm sgn}(\epsilon_\nu^R(z))={\rm sgn}(\epsilon_\nu(z)+
\epsilon_\nu(z^*))$ changes along the path connecting Euclidean
and Minkowski spaces. From the discussion in the previous section
it should be clear that such a change is most likely to
happen for the valence quark orbit. All other states possess
a sizable $|\epsilon_\nu^R|$ which is only slightly affected by
the $\omega$--meson. Let us for the moment consider the case with
baryon number $B=1$. The corresponding field configurations
usually admit one distinct quark level which is referred to
as the valence quark orbit. The static energy corresponding to
a configuration where this state is occupied is hence given by
\be
E&=&E_0+N_C\eta_{\rm val}\bar\epsilon_{\rm val}
\label{ee0v1} \\
&=&-\frac{N_C}{2}\sum_{\nu\ne{\rm val}}\bar\epsilon_\nu
+N_C\left(\eta_{\rm val}-\frac{1}{2}\right)\bar\epsilon_{\rm val}.
\label{ee0v2}
\ee
As already mentioned above, the states other than the valence state
are almost unaffected by the $\omega$--field. In particular only
${\rm sgn}(\epsilon_{\rm val}^R)$ might vary when performing the
analytic continuation. According to (\ref{defb}) the restriction to
$B=1$ demands for the valence quark occupation number
\be
\eta_{\rm val}=\frac{1}{2}\left(1+
{\rm sgn}(\epsilon_{\rm val}^R)\right).
\label{etaval}
\ee
Substituting this result into eqn (\ref{ee0v2}) yields
\be
E=\frac{N_C}{2}\left(\epsilon_{\rm val}-
\sum_{\nu\ne{\rm val}}\bar\epsilon_\nu\right)
\label{ee0v3}
\ee
which turns out to be analytic in $z$ in account on the above assertions
on ${\rm sgn}(\epsilon_{\nu\ne{\rm val}}^R)$.

The situation gets more involved when the fermion determinant is
regularized.  As already mentioned in the introduction the proper--time
regularization requires to treat real and imaginary parts (\ref{arai})
separately. This leads to distinct treatments of $\epsilon_\nu(z)$
and $\epsilon_\nu(z)^*$. In turn the energy functional will depend on
$z$ and its complex conjugate, {\it i.e.} $E=E(z,z^*)$. As only the
vacuum part $E_0$ undergoes regularization the transition from eqn
(\ref{ee0v1}) to eqn (\ref{ee0v2}) is, in general, prohibited. Thus there
are (at least) two points which may cause the regularized energy
functional to be non--analytic in $z$. Let us proceed step by step
in deriving the regularized energy functional to see how this comes
about.

At this point we wish to apply the proper--time regularization scheme
to the real part of the fermion determinant
\be
{\cal A}_R=\frac{1}{2}\left\{{\rm Tr}\ {\rm log}
\Big(-\partial_\tau+h(z)\Big)+
{\rm Tr}\ {\rm log}
\Big(\partial_\tau+h^{\dag}(z)\Big)\right\}
\label{arr1}
\ee
The functional trace is carried out
by summing over the eigen\-values\footnote{$\Omega_n=(2n+1)\pi/T$ are
the Matsubara frequencies. The fermionic character of the quarks requires
the associated eigenfunctions to assume anti--periodic boundary
conditions on the Euclidean time interval $T$.} $\Omega_n$ and
$\epsilon_\nu(z)$ corresponding to the operators $\partial_\tau$
and $h(z)$, respectively. As we intent to extract $E_0^R$
from the expression (\ref{arr1}) we need to consider the limit
$T\rightarrow\infty$.  Then the sum over $\Omega_n$ becomes a
spectral integral
\be
{\cal A}_R&=&\frac{N_C}{2}T\int_{-\infty}^\infty\frac{d\upsilon}{2\pi}
\sum_\nu\left\{{\rm log}\left(i\upsilon+\epsilon_\nu\right)
+{\rm log}\left(-i\upsilon+\epsilon_\nu^*\right)\right\}
\nonumber \\
&=&
\frac{N_C}{2}T\int_{-\infty}^\infty\frac{d\upsilon}{2\pi}\sum_\nu
{\rm log}\left\{\left|i\upsilon+\epsilon_\nu\right|^2\right\}.
\label{arr2}
\ee
To this expression the proper--time prescription may be applied
because the argument of the logarithm is positive definite
\be
{\cal A}_R=-\frac{N_C}{2}T\int_{-\infty}^\infty\frac{d\upsilon}{2\pi}
\sum_\nu\int_{1/\Lambda^2}^\infty \frac{ds}{s}\
{\rm exp}\left\{-s\left[\upsilon^2+\left|\epsilon_\nu(z)\right|^2
+2\upsilon\epsilon_\nu^I(z)\right]\right\}
\label{specint}
\ee
Here it has to be stressed that this spectral integral cannot be
obtained from the original definition of the proper--time
regularization for the real part (\ref{arreg}). Substituting
(\ref{edirac}) and (\ref{hz}) yields for (\ref{arreg})
\be
{\cal A}_R=
-\frac{1}{2}{\rm Tr}\ \int_{1/\Lambda^2}^\infty \frac{ds}{s}\
{\rm exp}\left\{-s\left[-\partial_\tau^2+h(z)h^{\dag}(z)
+\left(h(z)-h^{\dag}(z)\right)\partial_\tau\right]\right\}
\label{arreg1}
\ee
because $[\partial_\tau,h(z)]=0$. The spectral integral (\ref{specint})
could only be obtained from (\ref{arreg1}) under the assumption that
$h$ and $h^{\dag}$ can be diagonalized simultaneously. At the
Euclidean point this assumption is violated at quadratic order in
$\omega_4$. Technically the difference arises from the fact that
the usual rules for the logarithm have been used in order to derive
(\ref{arr2}). However, the original definition (\ref{arreg}) of the
proper--time regularization of ${\cal A}_R$ does not maintain these
rules for finite $\Lambda$ because the regularization curbs the Hilbert
space. Stated otherwise: Different regularization prescriptions induce
errors in the energy functional at order $\omega_4^2$. It should
furthermore be stressed that in deriving eqn (\ref{arr2}) one has
treated infinite quantities. This is avoided when the proper--time
prescription is imposed at the operator level (\ref{arreg},
\ref{arreg1}). In section 5 we will argue from the point of the
numerical results that (\ref{arreg1}) represents the more appropriate
starting point. Nevertheless let us continue with the investigation
of the expression (\ref{specint}) in particular to later on compare
the various approaches for treating the $\omega$--meson in the NJL model.

This spectral integral (\ref{specint}) converges absolutely permitting
in particular to shift the integration variable
$\upsilon\rightarrow\upsilon-\epsilon_\nu^I(z)$
\be
{\cal A}_R=-\frac{N_C}{2}T\int_{-\infty}^\infty\frac{d\upsilon}{2\pi}
\sum_\nu\int_{1/\Lambda^2}^\infty \frac{ds}{s}\
{\rm exp}\left\{-s\left[\upsilon^2+\left(\epsilon_\nu^R(z)\right)^2
\right]\right\}.
\label{specint1}
\ee
Performing the Gaussian integral allows one to extract
(${\cal A}_R\rightarrow-TE_0^R$)
\be
E_0^R(z,z^*)=\frac{N_C}{2}\sum_\nu\int_{1/\Lambda^2}^\infty
\frac{ds}{\sqrt{4\pi s^3}}\
{\rm exp}\left\{-s\left[\frac{\epsilon_\nu(z)+\epsilon_\nu(z^*)}{2}
\right]^2\right\}
\label{e0r}
\ee
which obviously depends on $z$ and its complex conjugate. In terms
of the eigenvalues $\Omega_n$ and $\epsilon_\nu(z)$ the
imaginary part of the action (\ref{arai}) reads
\be
{\cal A}_I=\frac{N_C}{2}\sum_{n,\nu}
{\rm log}\frac{i\Omega_n-\epsilon_\nu(z)}
{i\Omega_n-\epsilon_\nu(z^*)}.
\label{ai1}
\ee
Now it is important to note that the associated spectral integral
\be
\int_{-\infty}^\infty \frac{d\upsilon}
{2\pi}{\rm log}\frac{i\upsilon-\epsilon_\nu(z)}
{i\upsilon-\epsilon_\nu(z^*)}
\label{intnotconv}
\ee
is not absolutely convergent. If it were one could easily eliminate
the imaginary parts $\epsilon_\nu^I(z,z^*)$ of the energy eigenvalues
from the integral (\ref{intnotconv}) by reversing the sign of
$\upsilon$ in the denominator and subsequently shifting $\upsilon$
by $\epsilon_\nu^I(z,z^*)$. We therefore need an extended definition
of ${\cal A}_I$. We will later on see that the principle value
prescription
\be
{\cal A}_I=\frac{N_C}{2}T\sum_\nu{\cal P}
\int_{-\infty}^\infty \frac{d\upsilon}{2\pi}
{\rm log}\frac{i\upsilon-\epsilon_\nu(z)}
{i\upsilon-\epsilon_\nu(z^*)}
\label{aipv}
\ee
yields the desired physical symmetries such as the current
field identity. The integral (\ref{aipv}) can now be performed
using standard means
\be
{\cal A}_I&=&\frac{-iN_C}{2}\sum_\nu\, T\
{\cal P}\int_{-\infty}^{\infty}\frac{d\upsilon}{2\pi}
\int_{-1}^1 d\lambda\,
\frac{\epsilon_\nu^I}
{i(\upsilon-\lambda\epsilon_\nu^I)-\epsilon_\nu^R}
\nonumber \\
&=&\frac{-iN_C}{2}\sum_\nu\, T\
\lim_{{\cal M}\to\infty}\int_{-{\cal M}}^{{\cal M}}
\frac{d\upsilon}{2\pi}
\int_{-1}^1 d\lambda\, \frac{\epsilon_\nu^I}
{i(\upsilon-\lambda\epsilon_\nu^I)-\epsilon_\nu^R}.
\label{aiinteg}
\ee
Next the shift in the integration variable
$\upsilon-\lambda\epsilon_\nu^I \rightarrow \upsilon$ is performed.
It can be shown that the associated shift in the boundaries
does not contribute as ${\cal M}\to\infty$. This yields
\be
{\cal A}_I=\frac{-iN_C}{2}\sum_\nu \int_{-1}^1 d\lambda\, T
\lim_{{\cal M}\to\infty}\int_{-{\cal M}}^{{\cal M}}
\frac{d\upsilon}{2\pi}
\frac{\epsilon_\nu^I} {i\upsilon-\epsilon_\nu^R}.
\label{aiinteg1}
\ee
Now the integral over the parameter $\lambda$ may be done.
Although ${\cal A}_I$ is finite in the principle value formulation, the proper
time regularization may be imposed  by expressing the integrand as a
parameter integral:
\be
\frac{1}{\upsilon^2+(\epsilon_\nu^R)^2}\to\int_{1/\Lambda^2}^\infty ds\
{\rm exp}\left\{-s\left(\upsilon^2+(\epsilon_\nu^R)^2\right)\right\}
\label{aiprti}
\ee
which does obviously not diverge as $\Lambda\rightarrow\infty$.
We may finally extract the imaginary part of the Euclidean energy
$E_0^I$ from ${\A_I}\rightarrow -iTE_0^I$
\be
E_0^I(z,z^*)&=&\frac{iN_C}{4}\sum_\nu
(\epsilon_\nu(z)-\epsilon_\nu(z^*)) {\rm sgn}
(\epsilon_\nu(z)+\epsilon_\nu(z^*))
\nonumber \\
&& \hskip2cm \times
\cases{1,&$\A_I\quad {\rm not}\ {\rm regularized}$\cr
{\cal N}_\nu(z,z^*), &$\A_I\quad {\rm regularized}$\cr}
\label{e0i}
\ee
where we again made explicit the dependence on the complex variable
$z$ and its complex conjugate. Furthermore
\be
{\cal N}_\nu(z,z^*) = \frac{1}{\sqrt\pi}\Gamma\left(\frac{1}{2},
\left(\frac{\epsilon_\nu(z)+\epsilon_\nu(z^*)}{2\Lambda}\right)^2\right)=
{\rm erfc}\left(\left|\frac{\epsilon_\nu(z)+\epsilon_\nu(z^*)}
{2\Lambda}\right|\right)
\label{vacoccno}
\ee
are the vacuum ``occupation numbers" in the proper time regularization
scheme. The upper case in eqn (\ref{e0i}) corresponds to the limit
$\Lambda\rightarrow\infty$. Previously \cite{al93,zu94} the principle
value description was not stated explicitly. Fortunately, due to the
$\lambda$ integration the asymmetries in the boundaries cancel. Thus
the final result remains unchanged.

Analogously to the studies in section 2 we have then considered the
total energy functional originating from the fermion determinant
\be
E_{\rm det}(z,z^*)=E_0^R(z,z^*)+iE_0^I(z,z^*)+
N_C\eta_{\rm val}\bar\epsilon_{\rm val}(z,z^*)
\label{edet}
\ee
in the complex $z$--plane. The imaginary part $E_0^I$ is taken in
its regularized form. The behavior of $E(z,z^*)$ corresponding
to the parametrization (\ref{zpar}) is displayed in figure 3.1.
The cuts which occur for $\delta=2$ stem from level crossings as
described in the previous section. Otherwise $E(z,z^*)$ exhibits
a smooth behavior on the angle $\varphi$. Although the valence
and sea parts are treated differently the above indicated
non--analyticity (see the discussion starting at eqn (\ref{ee0v1})) due
to the change of ${\rm sgn}\left(\epsilon_{\rm val}^R(z,z^*)\right)$
is mitigated. This is a special feature of the proper--time
regularization which contains the equivalence between the limits
$\epsilon^R_\nu\rightarrow0$ and $\Lambda\rightarrow\infty$ in eqns
(\ref{e0r},\ref{e0i}). This is intuitively clear because these
regulator functions only depend on the dimensionless ratios
$\epsilon_\nu^R/\Lambda$. The smooth behavior of $E(z,z^*)$ in the
complex $z$--plane for the physically relevant radius $\delta\approx1$
suggests that $E(z,z^*)$ resembles an analytic function. This, of
course, would be somewhat astonishing because $z^*$ cannot be
expressed as a power series in $z$ without trading in an essential
singularity. As $E(z,z^*)$ depends on both $z$ and $z^*$ this
essential singularity should carry through to the energy functional.
In order to clarify the analytic properties of the energy functional
(\ref{edet}) we have also examined the corresponding Laurent series
\be
\tilde E_{\rm det}(z):=\sum_{n=-\infty}^\infty c_{{\rm det},n}(z_0)
(z-z_0)^n.
\label{lauredet}
\ee
Again we considered expansion centers located at the origin of
the complex plane ($z_0=0$) and the Euclidean point ($z_0=i$).
The results are summarized in table 3.1.

\begin{table}
\caption{Coefficients $c_{{\rm det},n}(z_0)$ in the Laurent series
for the contribution of the fermion determinant to the total energy
(\protect\ref{cdet}). These coefficients are measured in GeV. Also
shown are the associated continuations to the Minkowski point
$E_{\rm M}=\tilde E_{\rm det}(z=1)$ as obtained from the
coefficients quoted.}
{}~
\newline
\centerline{\tenrm\smalllineskip
\begin{tabular}{|c|c c c c|c c c c|}
\hline
 &\multicolumn{4}{c|}{$z_0=0$} & \multicolumn{4}{c|}{$z_0=i$} \\
$\delta$ & 0.1 & 1.0 & 1.2 & 1.5
& 0.1 & 1.0 & 1.2 & 1.5\\
\hline
$n=-4$ & 0.000     &-0.007          & 0.036     &-0.029
& 0.000         &-0.008          & 0.061-0.085$i$ & 0.031+0.043$i$ \\
$n=-3$ & 0.000     &-0.006          &-0.014     & 0.135
& 0.000         &-0.004+0.032$i$ & 0.066-0.013$i$ & 0.207+0.245$i$ \\
$n=-2$ & 0.000     &-0.064          &-0.136     &-0.260
& 0.000         &-0.045+0.012$i$ &-0.051-0.098$i$ &-0.111-0.192$i$\\
$n=-1$ & 0.000     &-0.005          & 0.069     & 0.330
& 0.003$i$      & 0.015+0.312$i$ & 0.039+0.370$i$ & 0.058+0.636$i$\\
$n=0$  & 1.208    & 1.348          & 1.425     & 1.572
&0.706+0.655$i$ & 0.922+0.654$i$ & 1.024+0.552$i$ & 1.202+0.481$i$\\
$n=1$  & 0.650    & 0.655          & 0.656     & 0.618
&0.689+1.164$i$ & 0.675+1.155$i$ & 0.649+1.078$i$ & 0.602+1.069$i$\\
$n=2$  & 0.740    & 0.742          & 0.744     & 0.717
&0.484-0.024$i$ & 0.528-0.013$i$ & 0.512-0.068$i$ & 0.510-0.048$i$\\
$n=3$  &-0.014    &-0.013          &-0.012     &-0.032
&0.008+0.164$i$ & 0.004+0.124$i$ &-0.028+0.088$i$ &-0.033+0.111$i$\\
$n=4$  & 0.042    & 0.041          & 0.039     & 0.028
&0.075-0.054$i$ & 0.023-0.003$i$ &-0.003-0.020$i$ & 0.009+0.004$i$\\
\hline
$E_{\rm M}(GeV)$ &2.627&2.700&2.673&2.782
&2.523+0.035$i$&2.712-0.036$i$&2.711+0.137$i$ & 2.715+0.029$i$ \\
\hline
\end{tabular}}
\end{table}
There are several observations to be made. First, we find non--vanishing
coefficients $c_{{\rm det},n}$ with $n<0$ even for parametrizations which
led to non--singular one--particle energies $\epsilon_\nu(z)$. Second,
these coefficients obviously depend on the path ($\partial A$) along
which the Cauchy integrals
\be
c_{{\rm det},n}(z_0)=\frac{1}{2\pi i}\oint_{\partial A}d\zeta
\frac{E_{\rm det}(\zeta,\zeta^*)}{\left(\zeta-z_0\right)^{n+1}}
\label{cdet}
\ee
are computed. The path $\partial A$ is again parametrized by
$\zeta=\delta{\rm exp}(i\varphi)$ with ($0\le\varphi\le2\pi$). Even
for small areas $A$ which include the expansion center ($z_0\in A$)
the coefficients $c_{{\rm det},n}$ do not converge. Especially the
coefficient of the constant term $c_{{\rm det},0}$ varies by about
20\% when decreasing $\delta$ from unity to $0.1$. For both values
of $\delta$ the one--particle energies were found to be analytic, see
section 2. Furthermore we extract from our numerical studies that (for
a given path $\partial A$) the coefficients $c_{{\rm det},n}$ decrease
only very slowly as $|n|$ gets larger. Thus we have accumulated (at
least) three arguments which support the conclusion that the Laurent
series (\ref{lauredet}) does not converge: The coefficients
$c_{{\rm det},n}$ with $n<0$ do not vanish, the radius of convergence
shrinks to zero, and the Minkowski extracted by analytic continuation
has a non--vanishing imaginary part. We have then
considered the continuation to the Minkowski point $E_{\rm M}=\tilde
E_{\rm det}(z=1)$. Again we find a dependence on $\delta$. This
dependence is somewhat weaker than in the case of the coefficients
$c_{{\rm det},n}$. We should, however, note that for the computation
of $E_{\rm M}$ only those coefficients which are displayed in table
3.1 were taken into consideration; {\it i.e.} $|n|\le4$. The reason
is that the numerical computation of the coefficients with larger $n$
contains significant errors for small $\delta$\footnote{The
$c_{{\rm det},n}$'s are of the generic form $I\times\delta^{-n}$,
{\it cf.} eqn (\ref{cpara}). Thus numerical errors in the integral
$I$ are multiplied for small $\delta$ and large $n$.}. A further
evidence for the non--existence of the series (\ref{lauredet}) can
be deduced from the fact that $E_M$ is not real when the expansion
is centered at the Euclidean point. We close our considerations on
the structure of $E_{\rm det}$ with the statement that the associated
Laurent series has vanishing radius of convergence. Stated more
drastically, the analytic continuation of $E_{\rm det}$ does not
exist. And in general
\be
E_{\rm det}(z,z^*)\not\equiv\tilde E_{\rm det}(z).
\label{nonanaly}
\ee
The Euclidean point resembles the only exception to this inequality.

The non--existence of an analytic continuation for $\delta\le1$ is
purely caused by the proper--time regularization of the Euclidean
action. As this regularization treats real and imaginary parts of the
one--particle energies differently it is interesting to consider other
regularization schemes which do not have this property. The most
suggestive approach would be to regularize the vacuum energy
(\ref{evacnonreg}) by employing a sharp cut--off and include all states
with $|\epsilon_\nu^R|\le\Lambda$ in the summation. We have also
studied the analytic properties of an energy functional defined in
this way. Indeed it yields a small but finite radius of convergence.
In this area of convergence the coefficients $c_{{\rm det},n}$ of
negative index $n<0$ vanish. Convergence is lost as one of the
real parts of the energy eigenvalues crosses $\pm\Lambda$ along
the path connecting Euclidean and Minkowski space. Since numerically
this radius of convergence is found to be significantly smaller
than unity (which is the physically relevant one) we will not
further pursue this treatment of the energy functional. Furthermore
a sharp cut--off regularization should be discarded because it does
not even preserve global gauge invariance.

\bigskip
\stepcounter{chapter}
\leftline{\large\it 4. Consistency Conditions for the Energy Functional}
\bigskip

In the last section we have learned that no analytic continuation of the
regularized energy functional exists. One therefore has to reside on
physical arguments, as {\it e.g.} gauge symmetry, in order to define
(or motivate) an energy functional in Minkowski space. This will be
the contents of the first part of this section. Actually three
distinct approaches to the Minkowski energy functional are known in
the literature. Historically the first to consider the $\omega$ meson
in the context of the NJL model were Watabe and Toki \cite{wa92}.
Secondly a different calculation performed by researchers at Bochum
University \cite{sch92,sch93} while the third approach was put forward
by the present authors \cite{al93,zu94}. In the second part of the
present section we will examine these approaches from two points of
view: (1) their relation to the Euclidean energy functional and (2)
their consistency with the conditions which will be derived in the
first part of this section.

A constant shift of the $\omega$--field $\omega(x)\rightarrow\omega(x)+
\omega_c$ is equivalent to a global gauge transformation and should
leave the fermion determinant unchanged, at least for topologically
trivial configurations. The non--regularized form of the fermion
determinant (\ref{dete}) can easily be seen to possess this property.
This is mostly a consequence of the principle value prescription for
the imaginary part of the action (\ref{aipv}).  At the Euclidean point
a constant shift in $\omega$ causes the imaginary part of the energy
eigenvalues to change accordingly
\be
\epsilon_\nu^I\rightarrow\epsilon_\nu^I+\omega_c
\qquad {\rm as}\qquad \omega(x)\rightarrow\omega(x)+\omega_c.
\label{shiftenu}
\ee
This straightforwardly transfers to that part of the energy
which stems from the fermion determinant
\be
E_0\rightarrow E_0-i\frac{N_C}{2}\omega_c\sum_\nu
{\rm sgn}\left(\epsilon_\nu^R\right)\qquad
{\rm and}\qquad
E_V\rightarrow E_V+iN_C\omega_c\sum_\nu\eta_\nu
{\rm sgn}\left(\epsilon_\nu^R\right)
\label{shifte1}
\ee
such that
\be
E_{\rm det}\rightarrow E_{\rm det}+iN_C B\omega_c.
\label{shiftedet}
\ee
This is just the desired result that shifting the $\omega$ field by
a constant changes the energy (associated with the fermion
determinant) by an amount proportional to the baryon number. This is
not surprising since a constant $\omega_c$ represents a chemical
potential for the quarks. Since
the unregularized form of the fermion determinant has been seen to
be analytic it is obvious that the relation (\ref{shiftedet}) can be
continued to Minkowski space. {\it I.e.} the Minkowski energy changes
by $N_C B\omega_c$. We therefore require a suitable definition of
the regularized energy functional in Minkowski space to transform
accordingly under $\omega(x)\rightarrow\omega(x)+\omega_c$. At this
point we also recognize the reason why a sharp cut--off regularization
in Minkowski space does not meet this requirement. In case $\omega_c$
shifts one of the eigenvalues across this sharp cut--off this special
state is dropped in the sum (\ref{evacnonreg}). Then the transformation
property (\ref{shiftedet}) is no longer valid. The same behavior can be
observed for other regularization schemes, as {\it e.g.} a three
dimensional momentum cut--off, which operate in Minkowski space. If one
wanted to implement the transformation relation (\ref{shiftedet})
for a three dimensional momentum cut--off in Minkowski space one
would inevitably be led to a regularization function which depended
on $\omega_c$. This feature is, of course, undesired. In
Euclidean space, however, only $\epsilon_\nu^R$ enters the
regularization condition. Thus the sharp cut--off or three momentum
regularization in Euclidean space is consistent with (\ref{shiftedet})
because the shift (\ref{shiftenu}) only effects the imaginary parts
$\epsilon_\nu^I$.

It should be remarked that the transformation property
(\ref{shiftedet}) holds in Skyrme type models as well. In those models
the coupling between $\omega$ and the chiral field is of the form
$N_C \omega_\mu B^\mu$ \cite{ad84}. Shifting the $\omega$ field by a
constant leads to the variation of the energy\footnote{Other couplings
appear as total derivatives of topologically trival objects and
thus do not contribute when shifting the $\omega$--field.}
\be
N_C \int d^3r\ \omega_c B_0=N_C \omega_c \int d^3r\ B_0=N_C \omega_c B.
\label{shiftskyrme}
\ee

In order to extract the baryon current one introduces an external
space--dependent gauge field $b_\mu(x)$ for the $U_V(1)$ symmetry.
The baryon current $B_\mu$ is then identified as the coupling linear
in $b_\mu(x)$, {\it i.e.}
\be
j_{I=0}^\mu=\frac{\delta{\cal A}[b]}
{\delta b_\mu(x)}\Big|_{b_\mu(x)=0} .
\label{defbmu}
\ee
The action ${\cal A}[b]$ is defined by replacing $\Vslash$
with $\Vslash+{b \hskip -0.4em /}$ in eqn (\ref{feract}). This
guarantees the local $U_V(1)$ symmetry. Under the functional integral
the vector meson fields may be shifted by an arbitrary amount (again
up to topological defects). This allows one to transfer the dependence
on the gauge field to the mesonic part of the action (\ref{mesact})
and straightforwardly obtain the baryon current
\be
j_{I=0}^\mu\propto\frac{1}{2G_2}{\rm tr}\left(V^\mu\right)=
\frac{1}{G_2}\omega^\mu
\label{calbmu}
\ee
{\it i.e.} the current field identity (\ref{i0current}) holds. In chiral
soliton models the identity (\ref{calbmu}) commonly is a consequence of
the stationary condition for the $\omega$--field. This condition is
obtained by extremizing the Minkowski energy functional. Thus we have
at hand a second consistency condition for the Minkowski energy
functional. For the static $\omega$--field this can also be interpreted
as the normalization of the $\omega$--field
\be
\int d^3r\ \omega(r)= G_2 B.
\label{normomega}
\ee
We will require the equation of motion for $\omega(r)$ to provide this
normalization. This normalization also has the important consequence
that the inclusion of the $\omega$ meson into the soliton calculation
does not necessarily lower the static energy although the number of
degrees of freedom is increased.

The first approach to incorporate the $\omega$ meson in the NJL soliton
calculation was carried out by Watabe and Toki\cite{wa92}. Their
calculation was completely performed in Minkowski space. In this
treatment the contribution of the fermion determinant to the energy
functional has been defined as
\be
E^{(1)}_{\rm det}=E^{(1)}_0+E^{(1)}_V
=\frac{N_C}{2}\sum_\nu
\int_{1/\Lambda^2}^\infty\frac{ds}{\sqrt{4\pi s^3}}\
{\rm exp}\left(-s\epsilon_\nu^2\right)
+N_C\eta_{\rm val}\epsilon_{\rm val}
\label{wte0}
\ee
wherein $\epsilon_\nu$ denote the eigenvalues of $h(z=1)$ (\ref{hz}).
Although this is formally the analytic continuation of (\ref{e0r}) it
has the disadvantage that the above stated consistency conditions
are not satisfied. {\it E.g.} the shift $\omega(x)\rightarrow
\omega(x)+\omega_c$ leads to $\epsilon_\nu\rightarrow\epsilon_\nu
+\omega_c$ and thus all powers of $\omega_c$ appear in (\ref{wte0})
after the shift. However, the requirement (\ref{shiftedet}) only allows
for a linear term. Furthermore as the derivative\footnote{This
derivative can, however, be identified as the regularized baryon number
carried by the polarized Dirac sea.}
\be
\sum_\mu\frac{\partial E^{(1)}_0}{\partial\epsilon_\mu}
=-\frac{N_C}{2}\sum_\mu{\rm sgn}\left(\epsilon_\mu\right)
{\rm erfc}\left(\left|\frac{\epsilon_\mu}{\Lambda}\right|\right)
\ne-\frac{N_C}{2}\sum_\mu{\rm sgn}\left(\epsilon_\mu\right)
\label{derwte0}
\ee
the normalization condition (\ref{normomega}) is not satisfied. Since
the analytic continuation of (\ref{edet}) does actually not exist
it is not surprising that the formal continuation does not meet the
consistency conditions. As these are not met by (\ref{wte0}) we
do ${\underline {\rm not}}$ consider it as a suitable definition for
the Minkowski energy functional.

Soon after this, Sch\"uren {\it et al}. \cite{sch92,sch93} presented
an approach which assumes as starting point the Euclidean action in
the proper--time regularization scheme (\ref{e0r},\ref{e0i}). In
order to reach the Minkowski point these authors ``interpreted" the
eigenvalues $\epsilon_\nu^\pm$ of the operators $h(z=\pm1)$ as the
continuations of $\epsilon_\nu(\pm i)$. Then
\be
E_{\rm det}^{(2)}=\frac{N_C}{2}\sum_\nu
\left\{\int_{1/\Lambda^2}^\infty
\frac{ds}{\sqrt{4\pi s^3}}{\rm exp}\left(-\frac{s}{4}
(\epsilon_\nu^++\epsilon_\nu^-)^2\right)
+\frac{1}{2}(\epsilon_\nu^+-\epsilon_\nu^-)
{\rm sgn}(\epsilon_\nu^++\epsilon_\nu^-)\right\}
\label{edetsch}
\ee
was defined as the contribution of the Dirac sea to the energy
functional\footnote{The imaginary part (\ref{e0i}) has been taken
in the unregularized version.}. Taking analyticity of the
one--particle energy eigenvalues for granted eqn (\ref{edetsch})
amounts to
\be
E_{\rm det}^{(2)}=E_{\rm det}(1,-1).
\label{e1m1}
\ee
Apparently this interpretation ${\underline{\rm cannot}}$ be
characterized as an analytical continuation. Advantageously the
definition (\ref{edetsch}) allows one to identify certain powers
of the $\omega$--field in the sence that $\epsilon_\nu^+$ is even
while $\epsilon_\nu^-$ is odd in agreement with $\epsilon_\nu^R$
and $\epsilon_\nu^I$, respectively. However, already the quadratic
order in $\omega$ (contained in $\epsilon_\nu^+$) of
$E_{\rm det}^{(2)}$ differs significantly from the expansion
in $\omega_4$ which starts from the expression
$\Dslash_E\Dslash_E^{\dag}$. This subject will be discussed at
length in section 5. Here it is merely important to keep in mind that
counting powers of $\omega$ in $\epsilon^\pm$ is (at least) suspicious;
most likely faulty. In contrast to the work of Watabe and
Toki \cite{wa92} the treatment of Sch\"uren {\it et al}. satisfies the
above derived consistency conditions. Shifting the $\omega$--field by
a constant implies $\epsilon_\nu^-\rightarrow\epsilon_\nu^-+\omega_c$
while $\epsilon_\nu^+$ remains unchanged. Substituting this
transformation prescription into eqn (\ref{edetsch}) immediately
provides the analytical continuation of (\ref{shiftedet}). The
normalization condition (\ref{normomega}) is verified by integrating
eqn (4.37) of ref.\cite{sch93}, which represents the equation of motion
in this approach, over the whole coordinate space.

The approach of the present authors to include the $\omega$ degree of
freedom into the NJL soliton calculations is based on an observation
made in section 2. When investigating the analytic structure of the
single quark eigenenergies it was found that in the physical relevant
region the Laurent series centered at the Euclidean point contained
only two significant terms, a constant and the expression linear in
the complex variable $z$, see tables 2.1 and 2.2. Transferring this
approximation to the total fermion determinant implies to identify
\be
E_{\rm det}(z)\approx E_0^R(i,-i)+ z E_0^I(i,-i)
\label{linapp}
\ee
because at the Euclidean point the real (imaginary) part contains only
even (odd) terms in $\omega_4$. This approximation is furthermore
supported by the fact that the ambiguities for regularizing ${\cal A}_R$
are of order $\omega_4^2$ (see the discussion proceeding eqn
(\ref{arreg1})). By construction, the approximation (\ref{linapp})
represents an analytic function of $z$ which can be continued to
Minkowski space yielding
\be
E_{\rm det}^{(3)}=E_0^R(i,-i) + E_0^I(i,-i).
\label{edetalk}
\ee
It should be stressed that in contrast to the two other approaches
the evaluation of $E_{\rm det}^{(3)}$ indeed requires the
diagonalization of a non--Hermitian Dirac Hamiltonian. The
one--particle energies which are evaluated in Euclidean space
enter this expression for the Minkowski energy functional whence
it is clear that the condition (\ref{shiftedet}) stemming from the
global gauge invariance is satisfied. The normalization condition
(\ref{normomega}) is proven to hold by integrating the equation
of motion for the $\omega$ profile function (Eqn (B.6) of
ref.\cite{zu94}) over the coordinate space\footnote{In ref.\cite{zu94}
we have considered the imaginary part of the action in the regularized
as well as in the unregularized form. The consistency conditions are
only satisfied for the unregularized form.}.

To conclude this section we would like to make some remarks on a
possible saddle point approximation to the functional integral in
Euclidean space. In this approximation the stationary conditions for
the profile functions are obtained by identifying the configuration
with the smallest real part of the Euclidean action. The reason is
that this configuration is assumed to dominate the functional integral.
The resulting equation of motion acquires the form
\be
0=\frac{\delta E_R(i,-i)}{\delta\varphi(r)}
\label{saddle}
\ee
where $E_R(i,-i)=E_0^R(i,-i)+E_m(i)$ also includes a part which
originates from the mesonic part of the action (\ref{mesact}) continued
to Euclidean space. Furthermore $\varphi$ denotes the meson field under
consideration. As $E_R(i,-i)$ contains only even powers of $\omega$
the saddle point equation (\ref{saddle}) possesses the undesired
solution $\omega\equiv0$. This solution does clearly not satisfy
the global gauge symmetry condition (\ref{shiftedet}). Thus the
saddle point condition has to be amended in order to comply with
the consistency conditions. Obviously the extremization of the
functional in which the fermion determinant is replaced by the
approximation (\ref{linapp}) at $z=1$ represents a suitable
extension of the saddle point equation. Although this replacement
is not unique it provides further justification for our treatment
of the $\omega$ meson.

\bigskip
\stepcounter{chapter}
\leftline{\large\it 5. Time Components up to Quadratic Order}
\bigskip

Up to this point the situation looks somewhat discouraging because
we have merely accumulated arguments why several approaches cannot
be considered correct. These approaches started on a complete
non--perturbative treatment of the $\omega$ meson in the NJL
soliton calculation. We have seen that then an exact analytic
continuation from Euclidean to Minkowski space is impossible for
the regularized module of the fermion determinant. In this section
we will thus return to a perturbative type of treatment for the
$\omega$--field and investigate up to which extent one can make
contact to the non--perturbative calculations discussed in the
previous sections.

We start off with the Euclidean Dirac Hamiltonian
$h_\Theta+i\omega_4$ and consider $\omega_4$ as a perturbation to
$h_\Theta$. Standard perturbation techniques\footnote{These techniques
assume a complete Hilbert space. However, regularization effectively
reduces this space.} may be employed for the energy eigenvalues to
illustrate the $\omega_4$ dependence
\be
\epsilon_\nu=\epsilon^0_\nu+\epsilon^1_\nu+\epsilon^2_\nu\ldots
\label{enuexp1}
\ee
with
\be
\epsilon^1_\nu=i\langle\nu|\omega_4|\nu\rangle \qquad
{\rm and}\qquad
\epsilon^2_\nu=-\sum_{\mu\ne\nu}\frac{\langle\nu|\omega_4|\mu\rangle
\langle\mu|\omega_4|\nu\rangle}{\epsilon^0_\nu-\epsilon^0_\mu}.
\label{enuexp2}
\ee
Here $\epsilon^0_\nu$ denote the eigenvalues of $h_\Theta$ and
$|\nu\rangle$ the corresponding eigenstates. The distinction between
real and imaginary parts is trivial
\be
\epsilon_\nu^R=\epsilon_\nu^0-\sum_{\mu\ne\nu}
\frac{\langle\nu|\omega_4|\mu\rangle
\langle\mu|\omega_4|\nu\rangle}{\epsilon^0_\nu-\epsilon^0_\mu}
\qquad {\rm and}\qquad
\epsilon_\nu^I=i\langle\nu|\omega_4|\nu\rangle.
\label{eripert}
\ee
\begin{table}
\caption{Perturbation for the one--particle energy eigenvalues
of the complex Dirac Hamiltonian in the $G^\pi=0^+$ channel.
All numbers are in units of the constituent quark mass $m$.}
{}~
\newline
\centerline{\tenrm\smalllineskip
\begin{tabular}{|c c c|c|c|}
\hline
\multicolumn{4}{|c|}{Perturbation} & \multicolumn{1}{c|}{Exact} \\
$\epsilon_\nu^0$  & $\epsilon_\nu^1$ & $\epsilon_\nu^2$
& total & $h(z=i)$ \\
\hline
-1.522 & 0.109$i$ &-0.016 &-1.537+0.109$i$ &-1.550+0.119$i$ \\
-1.293 & 0.093$i$ &-0.026 &-1.319+0.093$i$ &-1.333+0.086$i$ \\
-1.118 & 0.069$i$ &-0.042 &-1.159+0.067$i$ &-1.150+0.042$i$ \\
-1.029 & 0.018$i$ &-0.022 &-1.051+0.018$i$ &-1.037+0.069$i$ \\
 0.382 & 0.595$i$ & 0.019 & 0.401+0.595$i$ & 0.397+0.605$i$ \\
 1.071 & 0.011$i$ & 0.006 & 1.072+0.011$i$ & 1.072+0.009$i$ \\
 1.246 & 0.034$i$ & 0.004 & 1.240+0.034$i$ & 1.250+0.029$i$ \\
 1.481 & 0.056$i$ & 0.009 & 1.491+0.056$i$ & 1.490+0.050$i$ \\
\hline
\end{tabular}}
\end{table}
In table 5.1 we summarize the numerical results of this
perturbative calculation for the low--lying states in the $G^\pi=0^+$
channel which also contains the valence quark level. Again we have
adopted the standard reference profile functions already used
to gain the results presented in sections 2 and 3. We observe
that the perturbation series seems to converge quickly and that
the second order ($\epsilon_\nu^2$) is already negligibly small.
Furthermore the comparison with the exact diagonalization of the
Euclidean Dirac Hamiltonian shows reasonable agreement at the level
of a few percent. We have observed a similar behavior for states
not listed in table 5.1. This corroborates our earlier assumptions
which led to the approximation (\ref{linapp}).

It is then intuitive to transfer this expansion to the complete
energy functional at the Euclidean point. This will then be an
analytic function in $\omega_4$ which can straighforwardly be
continued to Minkowski space via $\omega_4\rightarrow -i\omega_0$.
Up to quadratic order in $\omega_4$ the energy functional
in Euclidean space is given by
\be
E_2^{\rm Eucl}[\Theta,\omega_4]&=&E_0[\Theta]
+iN_C\eta_{\rm val}\langle{\rm val}|\omega_4|{\rm val}\rangle
-i\frac{N_C}{2}\sum_\nu{\rm sgn}(\epsilon^0_\nu)
{\rm erfc}\left(\left|\frac{\epsilon^0_\nu}{\Lambda}\right|\right)
\langle\nu|\omega_4|\nu\rangle
\label{ethom} \\
&&\hspace{-2cm}-N_C\eta_{\rm val}\sum_{\nu\ne{\rm val}}
\frac{\left|\langle\nu|\omega_4|{\rm val}\rangle\right|^2}
{\epsilon^0_{\rm val}-\epsilon^0_\nu}+\frac{N_C}{2}
\sum_{\nu\mu}\tilde f\left(\epsilon^0_\nu,\epsilon^0_\mu;\Lambda\right)
\left|\langle\nu|\omega_4|\mu\rangle\right|^2
+\frac{4\pi}{3}\frac{m_\omega^2f_\pi^2}{m^2}\int dr r^2 \omega_4^2.
\nonumber
\ee
Here $E_0[\Theta]$ refers to the energy functional associated with
the Hermitian part of the Dirac Hamiltonian $h_\Theta$. Also the
subtraction for the trivial vacuum configuration $\Theta\equiv0$ is
contained in $E_0[\Theta]$. For the valence quark contributions
(proportional to $\eta_{\rm val}$) the application of (\ref{enuexp1})
and (\ref{enuexp2}) is straightforward. The corresponding
calculation for the Dirac sea, which yields the regulator function
\be
\tilde f\left(\epsilon_\nu,\epsilon_\mu;\Lambda\right)=\cases{
\frac{1}{2}\frac{{\rm sgn}(\epsilon_\nu)
{\rm erfc}\left(\left|\frac{\epsilon_\nu}{\Lambda}\right|\right)
-{\rm sgn}(\epsilon_\mu)
{\rm erfc}\left(\left|\frac{\epsilon_\mu}{\Lambda}\right|\right)}
{\epsilon_\nu-\epsilon_\mu}, & $\epsilon_\nu\ne\epsilon_\mu$\cr
\hspace{2cm}0, & $\epsilon_\nu=\epsilon_\mu$},
\label{fregal}
\ee
may be found in the appendix. Here it is important to remark that
this regulator function has vanishing diagonal elements
($\tilde f\left(\epsilon_\nu,\epsilon_\nu;\Lambda\right)=0$)
as a consequence of the second order perturbation formula
(\ref{enuexp2}). However, the limit $\epsilon_\mu\rightarrow
\epsilon_\nu$ is ${\underline {\rm not}}$ assumed smoothly. The last
term in eqn (\ref{ethom}) stems from (\ref{mesact}) with the gradient
expansion result \cite{eb86} $G_2=3m^2/2m_\omega^2f_\pi^2$
substituted\footnote{Here we do not include the effect of
$\pi-a_1$ mixing.}. In case one wants to consider the model with the
imaginary part of the action not being regularized the complementary
error function in eqn (\ref{ethom}) has to be replaced by unity. In
that case the consistency condition (\ref{shiftedet}) is satisfied.
This is easy to see since for a constant $\omega$--field
$\langle\nu|\omega_c|\mu\rangle=\omega_c\delta_{\nu\mu}$. As already
mentioned the diagonal elements do not contribute at quadratic
and higher order\footnote{This is due to the fact that the perturbation
of the wave--function is orthogonal to the unperturbed state in
Rayley--Schr\"odinger perturbation theory.} hence the whole change
under the constant shift is proportional to the baryon number. From
the numerical results listed in table 5.2 we find that the expansion
(\ref{ethom}) is reliable at the 3 percent level although the
contribution from the Dirac sea quadratic in $\omega_4$ is
unexpectedly sizable.
\begin{table}
\caption{Perturbation for the energy corresponding to the
fermion determinant at the Euclidean point. All numbers are
in MeV.}
{}~
\newline
\centerline{\tenrm\smalllineskip
\begin{tabular}{|c|c c c|c|c|}
\hline
&\multicolumn{4}{|c|}{Perturbation} & \multicolumn{1}{c|}{Exact} \\
&$0^{\rm th}$  & $1^{\rm st}$ & $2^{\rm nd}$
& $E_2^{\rm Eucl}$ & $E(z=i,z^*=-i)$ \\
\hline
$E_V$     & 458 & 714$i$ &  19 &477+714$i$ &476+726$i$ \\
$E_0+E_V$ &1261 & 681$i$ &-553 &724+681$i$ &705+692$i$ \\
\hline
\end{tabular}}
\end{table}

Although these results are encouraging we will now argue that
$\tilde f\left(\epsilon_\nu,\epsilon_\mu;\Lambda\right)$ in eqn
(\ref{fregal}) is not the correct regularization function at quadratic
order. We have already remarked the unpleasant feature that
$\tilde f\left(\epsilon_\nu,\epsilon_\mu;\Lambda\right)$ is
discontinuous as $\epsilon_\mu\rightarrow\epsilon_\nu$. A second
disadvantage of $\tilde f$ relates to the Minkowski energy and the
resulting equation of motion for $\omega_0$
\be
2\int dr^\prime r^{\prime2}
\tilde{\cal K}(r,r^\prime)\omega_0(r^\prime)
&=&\eta_{\rm val}\int d\Omega
\Psi^{\dag}_{\rm val}(\mbox{\boldmath $r$})
\Psi_{\rm val}(\mbox{\boldmath $r$})
\nonumber \\ &&
-\frac{1}{2}\sum_\nu{\rm sgn}\left(\epsilon^0_\nu\right)
{\rm erfc}\left(\left|\frac{\epsilon^0_\nu}{\Lambda}\right|\right)
\int d\Omega \Psi^{\dag}_{\rm val}(\mbox{\boldmath $r$})
\Psi_{\rm val}(\mbox{\boldmath $r$}).
\label{eqmomega}
\ee
Here we have defined the bilocal kernel
\be
\tilde{\cal K}\left(r,r^\prime\right)&=&
\frac{4\pi}{3}\frac{m_\omega^2f_\pi^2}{m^2}
\frac{\delta(r-r^\prime)}{r^2}
\nonumber \\ &&
+N_C\eta_{\rm val}\sum_{\nu\ne{\rm val}}
\int d\Omega \int d\Omega^\prime
\Psi_\nu^{\dag}(\mbox{\boldmath $r^\prime$})
\Psi_{\rm val}(\mbox{\boldmath $r^\prime$})
\frac{1}{\epsilon_\nu^0-\epsilon_{\rm val}^0}
\Psi_{\rm val}^{\dag}(\mbox{\boldmath $r$})
\Psi_\nu(\mbox{\boldmath $r$})
\nonumber \\ &&
+\frac{N_C}{2}\sum_{\nu\mu}
\int d\Omega \int d\Omega^\prime
\Psi_\nu^{\dag}(\mbox{\boldmath $r^\prime$})
\Psi_\mu(\mbox{\boldmath $r^\prime$})
\tilde f\left(\epsilon^0_\nu,\epsilon^0_\mu;\Lambda\right)
\Psi_\mu^{\dag}(\mbox{\boldmath $r$})
\Psi_\nu(\mbox{\boldmath $r$}).
\label{kern}
\ee
Since $\Psi_\nu(\mbox{\boldmath $r$})=\langle\mbox{\boldmath $r$}|
\nu\rangle$ denote the coordinate representations of the
eigenstates $|\nu\rangle$ of $h_\Theta$ this
kernel functionally depends on the chiral angle $\Theta$. Obviously
$\tilde{\cal K}(r,r^\prime)$ is symmetric in its arguments. The
{\it RHS} of the equation of motion (\ref{eqmomega}) is, of course,
nothing but the quark baryon density in the regularized
form\footnote{For our test profiles the question of regularizing the
baryon number or not plays a minor role since numerically the
regularized baryon number comes out to be 0.96.}. Upon integration
of this equation one easily verifies the normalization condition
(\ref{normomega}). Here the important ingredient is again the fact
that the regulator functions $\tilde f$ has vanishing diagonal
elements. By the virial theorem one can then eliminate the linear
term in the Minkowski energy functional for profiles which satisfy
(\ref{eqmomega})
\be
E_2^{\rm Mink}[\Theta,\omega_0]&=&E_0[\Theta]
+\frac{1}{2}\int dr r^2 \int dr^\prime r^{\prime2}
\tilde{\cal K}\left(r,r^\prime\right)\omega_0(r)\omega_0(r^\prime)
\label{evir}
\ee
Only in case that all eigenvalues of this kernel are positive the
Minkowski energy is positive definite. For the numerical treatment
we have discretized the radial coordinate leaving this kernel as a
matrix $\tilde{\cal K}(r,r^\prime)\rightarrow\tilde{\cal K}_{ij}$.
For various test profiles $\Theta$ we have found that this matrix
possesses ${\underline {\rm negative}}$ eigenvalues. This can easily
be understood by considering $\tilde f$ in eqn (\ref{fregal}).
The dominant contribution comes from states with $\epsilon_\nu\approx
\epsilon_\mu$. Assuming $\epsilon_\nu>\epsilon_\mu$ (without loss of
generality since $\tilde f$ is symmetric) we see that the denominator
is positive. On the other hand the numerator is negative because
the complementary error function decreases monotonically with the
argument. Thus $\tilde f$ may assume large negative values. Stated
differently, the fact that the Minkowski energy functional is not
positive definite for profiles satisfying the stationary condition
(\ref{eqmomega}) is a consequence of the discontinuity of
$\tilde f\left(\epsilon_\nu,\epsilon_\mu;\Lambda\right)$
as $\epsilon_\mu\rightarrow\epsilon_\nu$.

We have thus seen that counting powers of the $\omega$--field in
the eigenvalues of the Hamiltonian $h_\Theta+i\omega_4$ leads to
an ill--defined expression for the real part of the fermion determinant
in the proper--time regularization (\ref{arreg}). We also conjecture
that the non--positiveness of the energy functional based on this
power counting represents the reason why the authors of
ref.\cite{sch92} were unable to find self--consistent solutions
for physically relevant parameters.

To resolve this problem let us next examine ${\cal A}_R$ in more
detail. In order to expand ${\cal A}_R$ in terms of $\omega_4$ one
starts off at the module
\be
\Dslash_E\Dslash_E^{\dag}=-\partial_\tau^2+h_\Theta^2+
2i\omega_4\partial_\tau+i\left[h_\Theta,\omega_4\right]
+\omega_4^2.
\label{ddagom4}
\ee
Performing subsequently an expansion of
\be
{\cal A}_R&=&\frac{1}{2}{\rm Tr}\ {\rm log}
\left(-\partial_\tau^2+h_\Theta^2+
2i\omega_4\partial_\tau+i\left[h_\Theta,\omega_4\right]
+\omega_4^2\right)
\label{arom4} \\
&\rightarrow&
-\frac{1}{2}{\rm Tr}\ \int_{1/\Lambda^2}^\infty \frac{ds}{s}\
{\rm exp}\left(-\partial_\tau^2+h_\Theta^2+
2i\omega_4\partial_\tau+i\left[h_\Theta,\omega_4\right]
+\omega_4^2\right)
\label{arom4reg}
\ee
corresponds to imposing the proper--time regularization scheme
at the operator level. As has already been explained in section 3
the regularized form (\ref{e0r}) follows from an intermediate
definition of the proper--time regularization and involves the
assumption of simultaneous diagonalizability of $h$ and $h^{\dag}$.
This is exactly the point where $\omega_4$--power counting of the
eigenvalues fails to provide the correct result for the expansion of
${\cal A}_R$. We therefore expect different results for the action
functional when adopting (\ref{arom4reg}) as starting point for
expanding in  $\omega_4$ rather than expanding the energy eigenvalues.
Note that the expression (\ref{ddagom4}) is still exact
in $\omega_4$, {\it i.e.} no expansion has been performed up to here.
The vacuum contribution to the real part of the fermion determinant is
then expanded up to quadratic order in $\omega_4$ by methods which have
previously been worked out in the context of quantizing the chiral
soliton \cite{re89} and has subsequently been extended to study
fluctuations off the chiral soliton \cite{we93}. We refer the
interested reader to the appendix where the calculation is performed.
The net result of this calculation is that the regulator function
$\tilde f$ defined in eqn (\ref{fregal}) has to substituted by
\be
f\left(\epsilon_\mu,\epsilon_\nu;\Lambda\right)=
\frac{1}{2}\frac{{\rm sgn}(\epsilon_\mu)
{\rm erfc}\left(\left|\frac{\epsilon_\mu}{\Lambda}\right|\right)
-{\rm sgn}(\epsilon_\nu)
{\rm erfc}\left(\left|\frac{\epsilon_\nu}{\Lambda}\right|\right)}
{\epsilon_\mu-\epsilon_\nu}
-\frac{\Lambda}{\sqrt{\pi}}
\frac{e^{-\left(\epsilon_\mu/\Lambda\right)^2}-
e^{-\left(\epsilon_\nu/\Lambda\right)^2}}
{\epsilon_\mu^2-\epsilon_\nu^2}.
\label{freg}
\ee
The corresponding replacements have also to be performed in the
equations (\ref{ethom}) and (\ref{kern}). The latter replacement
defines the bilocal kernel ${\cal K}$ which then enters the new
equation of motion and energy functionals obtained by the
corresponding substitutions in equations (\ref{eqmomega}) and
(\ref{evir}), respectively. Apparently these two regulator functions
($\tilde f$ and $f$) only agree in the infinite cut--off limit,
$\Lambda\rightarrow\infty$. This is not astonishing because in this
limit the introduction of the Hermitian conjugate of the Dirac operator
is not mandatory in order to (formally) evaluate the fermion determinant.
It is also noteworthy that the regulator function $f$ (\ref{freg}) is
identical to the one which appears in the moment of inertia when the
semi--classical cranking approach is applied to quantize the chiral
soliton \cite{re89}. This is intuitively clear because for that problem
the perturbation (proportional to the isospin generators) in the
Dirac Hamiltonian is also static and anti--Hermitian. In that context
it has been shown that the limit
\be
\lim_{\epsilon_\mu\to\epsilon_\nu}
f\left(\epsilon_\mu,\epsilon_\nu;\Lambda\right)=0
\label{limf}
\ee
is assumed continuously (see chapter 5 in ref.\cite{re89}). Numerically
we have also verified that the associated kernel ${\cal K}$ possesses
positive eigenvalues only. Thus the corresponding energy functional
is positive definite for $\omega$ profiles which statisfy the modified
equation of motion (${\cal K}$ inserted in eqn (\ref{eqmomega})).
Due to the limit (\ref{limf}) it is clear that the Minkowski energy
functional with $\tilde f$ replaced by $f$ satisfies the consistency
conditions derived in section 4. For our standard test profile we
find that the second order in $\omega_0$ contributes about -129MeV to
the Minkowski energy. This is significantly smaller (in magnitude) than
for the regularization function $\tilde f$ (see table 5.2). Thus the
numerically results put further doubts on the validity of counting
$\omega_4$ powers in the energy eigenvalues. On the other hand,
employing the correct regulator function $f$ not only satisfies our
consistency conditions but also provides a quickly converging series.
The zeroth and first order contributions may be read off from table 5.2.
Thus we have, for the first time, available a well suited Minkowski
energy functional for the NJL model which includes the $\omega$ meson.
For completeness and later reference we list it here again although
the (formal) deviation from (\ref{ethom}) is only minor
\be
E^{\rm Mink}[\Theta,\omega_0]&=&E_0[\Theta]
+N_C\eta_{\rm val}\langle{\rm val}|\omega_0|{\rm val}\rangle
-\frac{N_C}{2}\sum_\nu{\rm sgn}(\epsilon^0_\nu)
{\rm erfc}\left(\left|\frac{\epsilon^0_\nu}{\Lambda}\right|\right)
\langle\nu|\omega_0|\nu\rangle
\label{emink} \\
&&\hspace{-2cm}-N_C\eta_{\rm val}\sum_{\nu\ne{\rm val}}
\frac{\left|\langle\nu|\omega_0|{\rm val}\rangle\right|^2}
{\epsilon^0_\nu-\epsilon^0_{\rm val}}-\frac{N_C}{2}
\sum_{\nu\mu}f\left(\epsilon^0_\nu,\epsilon^0_\mu;\Lambda\right)
\left|\langle\nu|\omega_0|\mu\rangle\right|^2
-\frac{4\pi}{3}\frac{m_\omega^2f_\pi^2}{m^2}\int dr r^2 \omega_0^2.
\nonumber
\ee
For $\omega$ profile which statisfy the associated stationary condition
this can be decomposed as
\be
E_2^{\rm Mink}[\Theta,\omega_0]=E_0[\Theta]+E_\omega[\Theta,\omega]
\label{eomega1}
\ee
where
\be
E_\omega[\Theta,\omega]=
\frac{1}{2}\int dr r^2 \int dr^\prime r^{\prime2}
{\cal K}\left(r,r^\prime\right)\omega_0(r)\omega_0(r^\prime)
\label{eomega2}
\ee
denotes the contribution due to the $\omega$ field.

In the next step one has to construct the self--consistent solution
which extremizes the functional (\ref{emink}). Unfortunately this is
not an easy task. The reason is that the kernel ${\cal K}$
functionally depends on the chiral angle $\Theta$. Thus the
stationary condition for $\Theta$ requires the functional derivative
of the wave--functions $\Psi_\nu(\mbox{\boldmath $r$})$ with respect
to $\Theta$. Such a calculation goes beyond the scope of the present
paper and we will pursue a different approach to find the energy
minimum. First, we introduce parameters $\{\chi\}$ to describe the
shape of the profile function $\Theta(r)$ and evaluate the corresponding
eigenvalues and --energies of $h_\Theta$. These are then functions of
the parameters $\{\chi\}$. In a second step this allows to construct
the kernel ${\cal K}$ as a function of these parameters as well. Via
the stationary condition for $\omega_0$, eqn (\ref{eqmomega}) with
$\tilde f$ replaced by $f$, also $\omega_0$ becomes a function of
these parameters. As already mentioned in the introduction
this stationary condition commonly represents a constraint rather
than an equation of motion. This fact furthermore justifies our
procedure to parametrize the chiral angle while solving exactly
for the $\omega$ profile. Technically this stationary condition is
an inhomogeneous integral equation which is solved by inverting the
discretized bilocal kernel ${\cal K}_{ij}$.  Finally the parameters
are tuned such that $E^{\rm Mink}$ acquires a minimum. Since the main
effect of the $\omega$ meson on the chiral angle is of repulsive
character it is intuitive to introduce a parameter which allows to
vary the spatial extension of $\Theta(r)$. In a first calculation we
therefore introduce the breathing mode parameter $\chi=\lambda$ via
\be
\Theta_\lambda(r)=\Theta_{\rm s.c.}(\lambda r)
\label{scalsc}
\ee
where $\Theta_{\rm s.c.}(r)$ denotes the self--consistent soliton
solution to the problem without the $\omega$ meson. The resulting
energy is displayed on left of figure 5.1 as a function of $\lambda$.
Secondly we have adopted the description \cite{ma86}
\be
\Theta_{R_m}(r)=-\pi\cases{1-\frac{r}{2R_r}, & $ r\le R_m $  \cr
\left(1-\frac{R_m}{2R_r}\right)\left(\frac{R_m}{r}\right)^2
\frac{1+m_\pi r}{1+m_\pi R_m}\
{\rm exp}\left(m_\pi(R_m-r)\right), & $ r\ge R_m $}
\label{rudi}
\ee
for the chiral angle. Since $R_r$ is chosen such that the derivative
of $\Theta$ is continuous the chiral angle only depends on $R_m$
parametrically. The results for the energy corresponding from the
parametrization (\ref{rudi}) may also be found in figure 5.1.
Actually these two parametrizations yield almost identical results.
For a constituent quark mass $m<400$MeV no local minimum is obtained
and the trivial minimum ($E=3m$) corresponding to three free quarks is
assumed. In the latter case the chiral angle gets very narrow while
the $\omega$--field tends to zero except within a vicinity of the
origin such that $|\int d^3r\ \omega|>0$ and
$|\int d^3r\ \omega^2|\rightarrow0$. As the constituent quark mass
is increased the local minimum gets more pronounced. In table 5.3 we
list the parameters describing the minimum as well as the resulting
total energy. There $\lambda^{\rm min}$ and $R_m^{\rm min}$ denote the
parameters associated with the energy minimum.
\begin{table}
\caption{Parameters describing the energy minimum as well
the corresponding total energy in Minkowski space, see eqn
(\protect\ref{emink}). Also shown is the contribution of the
$\omega$ field (\protect\ref{eomega2}).}
{}~
\newline
\centerline{\tenrm\smalllineskip
\begin{tabular}{|c|c c c|c c c|}
\hline
$m$ (MeV)& ~$\lambda^{\rm min}$~ & ~$E^{\rm Mink}$ (MeV)~
& ~$E_\omega$ (MeV)~
& ~$R_m^{\rm min}$ (fm)~ & ~$E^{\rm Mink}$ (MeV)~ &
{}~$E_\omega$ (MeV)~ \\
\hline
400 & 0.852 & 1586 & 321 & 0.713 & 1590 & 317 \\
450 & 0.776 & 1685 & 380 & 0.775 & 1685 & 378 \\
500 & 0.729 & 1769 & 428 & 0.830 & 1768 & 419 \\
\hline
\end{tabular}}
\end{table}
As already mentioned after eqn (\ref{normomega}) the increase of the
static energy due to the inclusion of the $\omega$ meson is no
surprise although the number of degrees of freedom has been enlarged.
The fact that at the minimum we always find $\lambda<1$ corroborates
the repulsive character of the $\omega$ meson. Now we are finally
able to mention that the test profiles, which we have been using
all the time, correspond to the parameterization
(\ref{scalsc}) for $m=400$MeV and $\lambda=0.86$, {\it i.e.} a
configuration which seems to be very close to the stationary solution.
Table 5.3 also contains the contribution of the $\omega$ meson to the
energy as defined in eqn (\ref{eomega2}). The fact that this quantity
is more sensible to the chosen parametrization than the total energy
indicates that a parameter ansatz for the soliton profiles is suitable
to compute the minimal energy but some refinements may be necessary
when studying other quantities, {\it e.g.} radii, magnetic moments,
etc. .  In figure 5.2 the profile functions which
minimize $E^{\rm Mink}$ in the parametrization (\ref{scalsc}) are
plotted.
In this figure the quark baryon density, $b\propto${\it RHS} of eqn
(\ref{eqmomega}), is scaled such that the spatial integrals over
$\omega/m$ and $b$ coincide. Obviously the $\omega$ profile is
enhanced over $b$ at large distances leading to a larger
isoscalar radius in agreement with assertion (\ref{i0radius}).
The fact that the presence of the $\omega$ meson causes the
soliton configuration to swell has also the consequences that the
axial charge becomes larger and that the valence quark gets bound
more strongly. These results were also obtained in the approach based
on the linear approximation (\ref{linapp}) \cite{zu94}. We have
furthermore considered fluctuations around the profile
$\Theta_{R_m^{\rm min}}(r)$ in order to obtain an improved solution.
The best we were able to achieve was a further decrease of the
total energy by about 5MeV, {\it i.e.} the parametrizations
(\ref{scalsc}) and (\ref{rudi}) allow for a good approximation to
exact minimum.

We should also note that we have performed the analogous calculations
using the regulator function $\tilde f$. In that case no local minimum
was obtained. This provides a further indication why the approach
of ref.\cite{sch92} did not lead to self--consistent solutions.

Here we will not further discuss the results for various observables
in this approach but rather make some remarks on the case when the
isovector mesons $\rho$ and $a_1$ are included. The previous formulas
apply as well, however, the Hermitian part of the Hamiltonian has to
be supplemented \cite{al92},
\be
h_\Theta&=&\mbox{\boldmath $\alpha$}\cdot\mbox{\boldmath $p$}
+m\beta\left({\rm cos}\Theta(r)+i\gamma_5
{\mbox{\boldmath $\tau$}}\cdot{\hat{\mbox{\boldmath $r$}}}
{\rm sin}\Theta(r)\right)
\nonumber \\*
&&+\frac{1}{2}(\mbox{\boldmath $\alpha $}\times
{\hat{\mbox{\boldmath $r$}}})\cdot{\mbox{\boldmath $\tau$}}G(r)
+\frac{1}{2}(\mbox{\boldmath $\sigma $}
\cdot{\hat{\mbox{\boldmath $r$}}})
(\mbox{\boldmath $\tau$}\cdot{\hat{\mbox{\boldmath $r$}}}) F(r)
+\frac{1}{2}(\mbox{\boldmath $\sigma$}\cdot
\mbox{\boldmath $\tau$})H(r).
\label{hvec}
\ee
The radial functions $G(r), F(r)$ and $H(r)$ denote the
vector ($V_\mu=\mbox{\boldmath $V$}_\mu\cdot\mbox{\boldmath $\tau$}/2$)
and axial vector ($A_\mu=\mbox{\boldmath $A$}_\mu\cdot\mbox{\boldmath
$\tau$}/2$)  meson profiles
\be
V_i^a=\epsilon_{aki}\hat r_k G(r) \qquad
A_i^a=\delta_{ia}H(r)+\hat r_i\hat r_a F(r)
\qquad (i,a=1,2,3).
\label{vecpro}
\ee
As the number of fields involved increases the parametrization of the
profile functions entering the Hermitian part of the Hamiltonian get
more ambiguous. Motivated by the breathing mode description one might,
in addition to (\ref{scalsc}), assume the {\it ans\"atze}
\be
\Phi_\lambda(r)=\lambda^n \Phi_{\rm s.c.}(\lambda r)\qquad
{\rm for}\quad \Phi=F,G,H.
\label{vecsc}
\ee
The choice $n=0$ would be analogous to (\ref{scalsc}) while
$n=1$ is suggested by the asymptotic form of the (axial) vector
meson fields in terms of the chiral angle.

Two important changes in comparison to the previous studies have
to be made in order to compute the total energy when the $\rho$ and
$a_1$ fields are included. First the mixing between the pions and
the axial vector meson field leads to a different relation between
the cut--off $\Lambda$ and the pion decay constant $f_\pi$ \cite{eb86}
\be
f_\pi^2=\frac{N_C m^2}{4\pi^2}
\Gamma\left(0,\left(\frac{m}{\Lambda}\right)^2\right)
\frac{m_\rho^2}{m_\rho^2+6m^2}.
\label{fpia1}
\ee
This increases the value for $\Lambda$ once a constituent quark mass
$m$ is given. Secondly the vector meson mass term becomes \cite{zu94}
\be
\frac{m_\rho^2}{2\pi}
\Gamma\left(0,\left(\frac{m}{\Lambda}\right)^2\right)
\int dr r^2\
\left[G^2(r)+\frac{1}{2}F^2(r)+F(r)H(r)+
\frac{3}{2}H^2(r)-2\omega_0^2(r)\right].
\label{emstatic}
\ee
We are now enabled to search for a local minimum using the
parametrization (\ref{vecsc}). In table 5.4 we present
values $\lambda^{\rm min}$ of the breathing mode variable which
lead to a local minima as well as the corresponding energies in
the interval $0\le n \le 2$. Here we use the constituent
quark mass $m=350$MeV and assume the imaginary part in the
unregularized from\footnote{When $\rho$ and $a_1$ fields are included
the regularized baryon number deviates strongly from unity \cite{zu94}
in contrast to the model with pseudoscalars only.}.
\begin{table}
\caption{The position ($\lambda^{\rm min}$) and the value
($E^{\rm Mink}$) of the local minimum of the energy when the
vector mesons are included. Also shown is the contribution of the
$\omega$ field (\protect\ref{eomega2}). The constituent quark mass
$m=350$MeV is assumed.}
{}~
\newline
\centerline{\tenrm\smalllineskip
\begin{tabular}{|c c|c c|}
\hline
$n$ & ~~$\lambda^{\rm min}$~~ & ~~$E^{\rm Mink}$ (MeV)~~
& ~~$E_\omega$ (MeV)~~ \\
\hline
0.00 & 0.850 & 1662 & 553 \\
0.25 & 0.800 & 1622 & 490 \\
0.50 & 0.745 & 1565 & 418 \\
0.75 & 0.680 & 1501 & 340 \\
1.00 & 0.630 & 1433 & 285 \\
1.25 & 0.600 & 1387 & 253 \\
1.50 & 0.603 & 1376 & 249 \\
1.75 & 0.625 & 1385 & 259 \\
2.00 & 0.655 & 1403 & 273 \\
\hline
\end{tabular}}
\end{table}
As a reminder we should mention that without the $\omega$ meson this
soliton energy is obtained to be $1010$MeV \cite{al92}. We observe that
a local minimal in the energy surface is obtained for $n\approx1.5$ and
$\lambda\approx0.58$. Upon further increase of $n$ the energy increases
rather slowly. In the same way the value of the scaling parameter
corresponding to the minimal energy, $\lambda^{\rm min}$, gets larger
again. The minimum of the total energy is also characterized by the
smallest contribution from the $\omega$ meson. It should be noted that
the minimal energy (1376MeV) is still 326MeV above the energy of the
trivial $B=1$ configuration where three (almost) free valence quark
orbits are occupied. Again we observe the repulsive character of the
$\omega$ meson reflected by the fact that $\lambda^{\rm min}<1$. In
contrast to the case with only the pseudoscalar fields this leads to a
somewhat smaller binding of the valence quarks. We find
$\epsilon_{\rm val}^0/m=-0.29$ to be compared with $-0.38$ in the model
without the $\omega$ \cite{al92}. This reduction seems to be due the
diminishing profile functions of the (axial) vector mesons.
A similar behavior has been obtained when extremizing the energy
functional involving the definition (\ref{edetalk}) in the case
that the imaginary has not been regularized \cite{zu94}. As this
prescription is also adopted here we recognize at least qualitative
similarities. The value for the axial charge of the nucleon
as obtained from the integral \cite{zu94}
\be
g_A=-\frac{2\pi}{3g_2}\int dr r^2\left[H(r)+\frac{1}{3}F(r)\right]
\label{ga}
\ee
increases from $0.27$ to $0.48$ by the inclusion of the
$\omega$ meson. This is, of course, a correction
into the right direction but nevertheless only about $1/3$ of the
empirical value extracted from neutron $\beta$--decay. It has been
argued that $1/N_C$ corrections may further improve on the result
for $g_a$ \cite{wa93}; these arguments have, however, to be taken
with some care in respect to the symmetries of the model \cite{al93a}.

As the main subject of the present paper is the investigation
of the $\omega$ meson in the context of the energy functional we
will not go into detail on the nucleon properties. Especially because
we only have a parametrical description of the stationary field
configuration rather than a self--consistent solution. The nucleon
properties are most likely more sensitive to the shape of the meson
profiles and the quark wave--functions than the soliton energy.
Nevertheless the construction of the self--consistent solution would be
an interesting path to pursue in particular in order to gain
information about the nucleon properties. In this context we also
wish to remark that the $\omega$ profile associated with the minimal
energy configuration ($n=1.5,\lambda^{\rm min}=0.598$) does not turn
out to be a smooth radial function but rather possesses small
fluctuations off a monotonously decreasing function. We expect an
exact solution to improve in this respect.

\bigskip
\stepcounter{chapter}
\leftline{\large\it 6. Summary and Conclusions}
\bigskip

In this paper we have performed a thorough study of the analytical
properties of a fermion determinant for the case that the $\omega$--field
is present in the static hedgehog configuration. This has
recently been a matter of dispute in the context of the soliton
in the NJL model with (axial) vector meson degrees of freedom.
The fact that this model requires regularization makes mandatory
the continuation forth and back from Minkowski to Euclidean space.
This continuation represents an element of the transformations
we have been considering in the complex $z$--plane defined by the
Dirac Hamiltonian (\ref{hz}). As originally the fermion determinant
is expressed in terms of the eigenvalues of this non--Hermitian
operator we first have investigated the behavior of the associated
eigenvalues in the $z$--plane. Thereby we have confirmed an
earlier result \cite{sch93tr} that for physically motivated field
configurations these eigenvalues are smooth functions along the
path connecting Euclidean and Minkowski spaces. Furthermore we
have demonstrated that the Laurent series of such an eigenvalue
or --function indeed reduces to a Taylor series. In this way the
analyticity of eigenvalues and --functions is verified. Such an
analytic behavior is not {\it a priori} clear because the
eigenvalues are roots of a polynomial of large degree. We have,
however, encountered a different kind of non--analyticity which
is caused by a level crossing. Such crossing may appear if the
$\omega$--field is sufficiently pronounced. Thus, in order to
attach physical significance to a given field configuration one
has to check that no such level crossing takes place along the
path from Euclidean to Minkowski spaces.

In the next step we have then examined the behavior of the full
fermion determinant in the complex $z$--plane. Formally analytical
one--particle eigenvalues yield an analytical determinant in the
unregularized form. However, regularization is unavoidable
and introduces (at least in the proper--time scheme) the
Hermitian conjugate of the Euclidean Dirac operator $\Dslash_E^{\dag}$.
As a consequence the real and imaginary parts of the one--particle
eigenvalues are treated differently when summing up for the fermion
determinant. This represents the main reason for the fermion
determinant to be non--analytical in $z$. As a matter of fact the
corresponding Laurent series (\ref{lauredet},\ref{cdet}) has
vanishing radius of convergence. We therefore conclude that the
analytical continuation of the Euclidean action functional to
Minkowski space does not exist. This makes a derivation of the
Minkowski energy functional by means of analytic continuation
impossible.

In order to nevertheless obtain a sensible definition of a Minkowski
energy functional we have then formulated two consistency conditions
to be satisfied by this functional. The first one says that the
energy changes by $N_CB\omega_c$ under a constant shift ($\omega_c$)
of the $\omega$--field as a consequence of global gauge invariance
($B$ is the baryon number.). Secondly, the current field identity for
the baryon current imposes a normalization on the $\omega$--field
(\ref{normomega}). We have then examined the three treatments known
in the literature with respect to these conditions. In this context
only two approaches (by Sch\"uren et al. \cite{sch92,sch93} and
the one by the present authors \cite{al93,zu94}) were shown to
be consistent. The reason is that these two approaches are motivated
by the Euclidean energy functional in contrast to the treatment
by Watabe and Toki \cite{wa92}.

In the last section we have then investigated the expansion of the
fermion determinant in powers of the $\omega$--field.
In particular we have examined the role of $\Dslash_E^{\dag}$. For this
purpose we have considered the $\omega$--field as a perturbation to
the Euclidean Dirac Hamiltonian and applied standard perturbation
techniques. We have then demonstrated that at second order the
bilocal kernel $\tilde {\cal K}$ in eqn (\ref{kern}) is not positive
definite putting some doubts on the validity on an approach which is
based on counting $\omega_4$ powers in the eigenvalues $\epsilon_\nu$.
The reason is that the corresponding regularization function
$\tilde f\left(\epsilon_\nu,\epsilon_\mu;\Lambda\right)$ does not
smoothly assume the limit $\tilde f\rightarrow0$ as
$\epsilon_\mu\rightarrow\epsilon_\nu$. A vanishing regularization
function for identical arguments is required in order to satisfy
the consistency conditions derived in section 4.
As a matter of fact the functional trace of the operator
$\Dslash_E\Dslash_E^{\dag}$ rather than the eigenvalues of the Dirac
Hamiltonian is required in order to evaluate the fermion determinant
in the proper--time regularization scheme. Stated otherwise, the
proper--time prescription has to be imposed at the operator level. This
point of view has already been adopted previously when performing the
semi--classical quantization for two \cite{re89} and three flavors
\cite{we92}.  We have then observed that imposing the proper--time
prescription at the operator level leads to a different expansion in
terms of the $\omega$--field for the action functional. From a
mathematical point of view there are two ways for understanding the
different results. First, in order to obtain the regularized form of
the energy functional (\ref{e0r}) from the regularized action
(\ref{arreg}) one has to make the inexact assumption that the
Euclidean Dirac Hamiltonian and its complex conjugate can be
diagonalized simultaneously. It is then also clear that both
expansion schemes yield identical results when the perturbation of
the Dirac Hamiltonian is static and Hermitian. We have verified this
for an expansion up to second order. The second explanation relies on
the fact that due to regularization the Hilbert space gets restricted
and thus the rules for handling logarithms are no longer valid.
Adopting this explanation it becomes also clear that in the limit
$\Lambda\rightarrow\infty$ (no regularization) the two expansion
schemes are again identical. In regard of the fact that the derivation
of the functional (\ref{arr2}) involves a non--converging quantity, the
results for the semi-classical quantization \cite{re89,we92}, the
smoothness of the regulator function $f$ (\ref{freg}) and the existence
of a local minimum in the energy surface (\ref{emink}) we consider the
imposition of the proper--time regularization prescription at the
operator level as the correct one. This conclusion finally prohibits
an expansion of the eigenvalues $\epsilon_\nu$ in order to obtain the
Minkowski energy functional in the presence of the $\omega$ meson. We
have also seen that a series based on the expansion of
$\Dslash_E\Dslash_E^{\dag}$ converges rather quickly (at least at
second order).

This conclusion has enforced us to investigate the corresponding
energy functional (\ref{emink}) in more detail. Employing a
parametrical description of the chiral angle we have constructed
the local minimum to this functional. Local minima exist for
constituent quark masses $m{_>\atop^\sim}400$MeV. We have great
confidence that the one--parameter {\it ans\"atze}
(\ref{scalsc},\ref{rudi}) describe the exact solution
quite well. For these solutions we have confirmed the repulsive
character of the $\omega$--field. When other vector meson fields
entering the Hermitian part of the Dirac Hamiltonian are included
as well, a parametrical description of the Dirac Hamiltonian is
more ambiguous. Nevertheless we have demonstrated that for the
constituent quark mass $m=350$MeV a solution exists. In particular,
the result that the valence quark level joins the Dirac vacuum, which
was previously obtained in calculations ignoring the $\omega$ meson
\cite{al92}, is affirmed. This is again mainly due to the repulsive
character of the isoscalar $\omega$.

Let us finally add a note on similar soliton models which include the
$\omega$ meson. In extended Skyrme models \cite{me87,ad84,re86,sch89}
or the chiral quark model \cite{co89} the $\omega$ meson appears only in
terms of explicit polynomials which renders the energy functional
analytic in the complex $z$--plane. More importantly these models
do actually not require regularization abandoning the need for
continuation forth and back from Minkowski to Euclidean spaces.

In order to point out an interesting task to pursue in future
we should mention that we have not yet solved the complete set
of stationary conditions in the presence of all meson fields.
In the present work we have only made plausible that such a
solution exists. However, a calculation aimed at the extremization
of the functional (\ref{emink}) with all meson profiles included,
which are compatible with the grand spin and parity symmetries
of the hedgehog {\it ansatz}, is highly desirable because we have
(for the first time) available a reasonable approximation
(\ref{emink}) to the energy functional of the NJL soliton model
including all relevant meson fields. This should represent a well
suited model for the description of static nucleon properties.

\bigskip
\nonumsection{Acknowledgement}
\bigskip
One of us (HW) gratefully acknowledges a fruitful conversation with
C. Sch\"uren, E. Ruiz Arriola and K. Goeke at the ${\rm ETC}^*$
workshop in Trento on {\it The Quark Structure of Baryons}, October
1993.

\appendix

\vskip2cm
\stepcounter{chapter}
\leftline{\large\it Appendix: Regularization for Quadratic Terms}
\bigskip

In this expansion we wish to describe the expansion of the vacuum
part of the Euclidean energy functional in terms of the isoscalar
vector $\omega$--meson. The
starting point is represented by the Euclidean Dirac operator
\be
i\beta\Dslash_E=-\partial_\tau-h=-\partial_\tau-h_\Theta-i\omega_4
\label{deapp}
\ee
with the Dirac Hamiltonian being decomposed into Hermitian ($h_\Theta$)
and anti--Hermitian ($i\omega_4$) pieces, see also eqn (\ref{decomph}).
In order to extract the vacuum part we will only consider the limit
of large Euclidean times $T\rightarrow\infty$. We actually need to expand
\be
{\cal A}_E={\rm Tr}\ {\rm log}\left(
-\partial_\tau-h_\Theta-i\omega_4\right)
\label{aeapp}
\ee
in powers of the $\omega_4$ field: ${\cal A}_E={\cal A}^0_E+
{\cal A}^1_E+{\cal A}^2_E+\ldots$. Here the superscripts
label the order in which the $\omega$ field appears.
The term linear in $\omega_4$ is straightforwardly obtained
to become purely imaginary
\be
{\cal A}^1_E=-i{\rm Tr}\left[\left(\partial_\tau+h_\Theta\right)^{-1}
\omega_4\right].
\label{ae11}
\ee
In the limit of large Euclidean times ($T\rightarrow\infty$) the
temporal part of the trace is replaced by a spectral integral
while the trace over the remaining degrees of freedom is carried
out using the eigenstates of $h_\Theta$ ({\it cf}. section 3). We
would like to remark that the corresponding eigenvalues $\epsilon^0_\nu$
are real. Then one easily finds
\be
{\cal A}^1_E=-iN_C\sum_\nu{\cal P}\int\frac{d\upsilon}{2\pi}
\left(i\upsilon+\epsilon^0_\nu\right)^{-1}
\langle\nu|\omega_4|\nu\rangle
=-iN_C\sum_\nu\int_{-\infty}^\infty\frac{d\upsilon}{2\pi}
\frac{\epsilon_\nu}{\upsilon^2+(\epsilon^0_\nu)^2}
\langle\nu|\omega_4|\nu\rangle.
\label{ae12}
\ee
Again one may introduce a finite proper--time cut--off $\Lambda$
by expressing the denominator in eqn (\ref{ae12}) in terms of a
parameter integral
\be
{\cal A}^1_E=-iN_C T\sum_\nu\epsilon^0_\nu\int_{-\infty}^\infty
\frac{d\upsilon}{2\pi} \int_{1/\Lambda^2}^\infty ds\ {\rm exp}
\left[-s\left(\upsilon^2+(\epsilon^0_\nu)^2\right)\right]
\langle\nu|\omega_4|\nu\rangle.
\label{ae13}
\ee
The spectral integral may now be performed which results in
\be
{\cal A}^1_E=\frac{-i}{2}N_C T\sum_\nu{\rm sgn}(\epsilon^0_\nu)
{\rm erfc}\left(\left|\frac{\epsilon^0_\nu}{\Lambda}\right|\right)
\langle\nu|\omega_4|\nu\rangle.
\label{ae14}
\ee
At this point is important to remark that this result can also
be gained by the application of standard perturbation techniques
to the expression for the Euclidean energy functional obtained in
section 3. In a first step one expands the functional (\ref{edet})
for $z=i$ and $z^*=-i$ in terms of the one--particle eigenenergies.
Secondly, standard perturbation theory is employed to make the
$\omega_4$ dependence explicit as described in eqns (\ref{enuexp1})
and (\ref{enuexp2}). Carrying out this procedure up to linear
order in $\omega_4$ exactly reproduces eqn (\ref{ae14}). The term
of quadratic order $\epsilon^2_\nu$ will be considered later.

Next we turn to the evaluation of ${\cal A}^2_E$. As the terms of
even powers in $\omega_4$ contribute to the real part of the
fermion determinant we may start the definition of the proper--time
regularization (\ref{arreg})
\be
{\cal A}_R=-\frac{1}{2}{\rm Tr}
\int_{1/\Lambda^2}^\infty\frac{ds}{s}\ {\rm exp}
\left(-s\Dslash_E\Dslash_E^{\dag}\right).
\label{ae21}
\ee
In the presence of the $\omega$ meson the operator
$\Dslash_E\Dslash_E^{\dag}$ is presented in eqn (\ref{ddagom4}).
We then apply techniques which have previously been developed for
the description of meson fluctuation off the soliton\cite{we93} to
this expression. As the $\omega$ field is assumed to be static
the expansion is also similar to the $1/N_C$ expansion performed
to extract the moment of inertia in the cranking approach\cite{re89}.
The non--vanishing contributions at quadratic order in $\omega_4$
are found to be
\be
{\cal A}_R^2&=&
\frac{1}{2}{\rm Tr}\int_{1/\Lambda^2}^\infty ds\
{\rm exp}\left[-s\left(h_\Theta^2-\partial_\tau^2\right)\right]
\omega_4^2
\label{ae22} \\ &&
-\frac{1}{2}{\rm Tr}\int_{1/\Lambda^2}^\infty ds s
\int_0^1dx\int_0^{1-x}dy\
{\rm exp}\left[-sx\left(h_\Theta^2-\partial_\tau^2\right)\right]
\left\{2i\omega_4\partial_\tau+
i\left[h_\Theta,\omega_4\right]\right\}
\nonumber \\
&&\hspace{0.2cm}\times
{\rm exp}\left[-s(1-x-y)\left(h_\Theta^2-\partial_\tau^2\right)\right]
\left\{2i\omega_4\partial_\tau+
i\left[h_\Theta,\omega_4\right]\right\}
{\rm exp}\left[-sy\left(h_\Theta^2-\partial_\tau^2\right)\right].
\nonumber
\ee
One of the Feynman parameter integrals can be carried out trivially due
to the cyclic properties of the trace. Furthermore the temporal part of
the trace turns into a spectral integral for $T\rightarrow\infty$.
\be
{\cal A}_R^2&=&
\frac{1}{2} T{\rm tr}\int_{1/\Lambda^2}^\infty ds\
\int_{-\infty}^\infty\frac{d\upsilon}{2\pi}
{\rm exp}\left[-s\left(h_\Theta^2+\upsilon^2\right)\right]
\omega_4^2
\label{ae23} \\ &&
-\frac{1}{4} T{\rm tr}\int_{1/\Lambda^2}^\infty ds s
\int_{-\infty}^\infty\frac{d\upsilon}{2\pi}\int_0^1d\alpha\
{\rm exp}\left[-s\alpha\left(h_\Theta^2+\upsilon^2\right)\right]
\left\{2\omega_4\upsilon+i\left[h_\Theta,\omega_4\right]\right\}
\nonumber \\
&&\hspace{3.0cm}\times
{\rm exp}\left[-s(1-\alpha)\left(h_\Theta^2+\upsilon^2\right)\right]
\left\{2i\omega_4\upsilon+i\left[h_\Theta,\omega_4\right]\right\}.
\nonumber
\ee
These integrals are Gaussian and may readily be carried out. The
remaining trace (${\rm tr})$ is computed using the eigenstates of
$h_\Theta$. These manipulations then result in
\be
{\cal A}_E^2&=&\frac{N_C}{2}T\sum_\nu\int_{1/\Lambda^2}^\infty
\sum_\nu\frac{ds}{\sqrt{4\pi s}}e^{-s(\epsilon^0_\nu)^2}
\langle\nu|\omega_4|\nu\rangle
\label{ae24} \\ &&
-\frac{N_C}{4}T\sum_{\nu\mu}
\int_{1/\Lambda^2}^\infty \frac{ds}{\sqrt{4\pi s^3}}
\frac{2+s(\epsilon^0_\nu-\epsilon^0_\mu)^2}
{(\epsilon_\nu^0)^2-(\epsilon_\mu^0)^2}
\left(e^{-s(\epsilon^0_\nu)^2}-e^{-s(\epsilon^0_\mu)^2}\right)
\langle\nu|\omega_4|\mu\rangle \langle\mu|\omega_4|\nu\rangle.
\nonumber
\ee
Making use of closure as well as of the invariance of the second
term under $\epsilon_\mu^0\leftrightarrow\epsilon^0_\nu$ this
expression may be rewritten as
\be
{\cal A}_E^2&=&\frac{N_C}{2}T\sum_{\mu\nu}
\int_{1/\Lambda^2}^\infty \frac{ds}{\sqrt{4\pi s}}
\langle\nu|\omega_4|\mu\rangle \langle\mu|\omega_4|\nu\rangle
\nonumber \\ && \hspace{1cm}\times
\left[\frac{\epsilon^0_\mu e^{-s(\epsilon^0_\mu)^2}
+\epsilon^0_\nu e^{-s(\epsilon^0_\nu)^2}}
{\epsilon^0_\mu+\epsilon^0_\nu}
+\frac{1}{s}\frac{e^{-s(\epsilon^0_\mu)^2}-e^{-s(\epsilon^0_\nu)^2}}
{(\epsilon^0_\mu)^2-(\epsilon^0_\nu)^2}\right]
\label{ae25}
\ee
from which $f\left(\epsilon_\mu,\epsilon_\nu;\Lambda\right)$ in
eqn (\ref{freg}) can be extracted after performing an integration
by parts in the last term.
Next the question arises whether the application of standard
perturbation techniques yields the same result. As already discussed
in the main part of this paper it does not! In order to see how this
comes about we start from the expression of the real part of the
action (\ref{specint1}) and perform the spectral integration
\be
{\cal A}_R=-\frac{N_C}{2}T\sum_\nu\int_{1/\Lambda^2}^\infty
\frac{ds}{\sqrt{4\pi s^3}}\
{\rm exp}\left(-s(\epsilon_\nu^R)^2\right).
\label{aee1}
\ee
Using the expansion (\ref{enuexp1}) for the energy eigenvalues
($\epsilon_\nu^R=\epsilon_\nu^0+\epsilon_\nu^2\ldots$)
\be
\tilde{\cal A}_E^2&=&N_C T \sum_\nu\int_{1/\Lambda^2}^\infty
\frac{ds}{\sqrt{4\pi s}}\
\epsilon_\nu^0\epsilon_\nu^2
{\rm exp}\left(-s(\epsilon_\nu^R)^2\right)
\label{aee1p}
\ee
and substituting the expression (\ref{enuexp2}) gives
\be
\tilde{\cal A}_E^2&=&-\frac{N_C}{2}T\sum_{\mu\ne\nu}
\frac{{\rm sgn}(\epsilon_\nu^0)}{\epsilon_\nu^0-\epsilon_\mu^0}
{\rm erfc}\left(\left|\frac{\epsilon_\nu^0}{\Lambda}\right|\right)
\langle\nu|\omega_4|\mu\rangle \langle\mu|\omega_4|\nu\rangle
\nonumber \\
&=&-\frac{N_C}{4}T\sum_{\mu\ne\nu}
\frac{{\rm sgn}(\epsilon_\nu^0)
{\rm erfc}\left(\left|\frac{\epsilon_\nu^0}{\Lambda}\right|\right)-
{\rm sgn}(\epsilon_\mu^0)
{\rm erfc}\left(\left|\frac{\epsilon_\mu^0}{\Lambda}\right|\right)}
{\epsilon_\nu^0-\epsilon_\mu^0}.
\label{aee2}
\ee
For the energy functional this leads to the expression $\tilde
f\left(\epsilon_\mu,\epsilon_\nu;\Lambda\right)$ given in eqn
(\ref{fregal}).

\vfil\eject

\centerline{\Large \bf Figure captions}

\vskip1cm

\centerline{\large \bf Figure 2.1}

\noindent
Left: The behavior of the valence quarks' eigenenergy in the complex
plane $z=\delta e^{i\varphi}$. The starting point $\varphi=0$ is
indicated. From there it continues anti--clockwise as $\varphi$
increases.  Right: The same for the state with the smallest (positive)
real part of the energy eigenvalue in the $G^\pi=2^+$ channel.

\vskip1cm
\centerline{\large \bf Figure 2.2}

\noindent
The behavior of the valence quarks' eigenfunction in the
complex plane in arbitrary units. $\varphi$ labels the phase
of the point in the complex plane $z=\delta e^{i\varphi}$.
Left: At $r=0$, only the upper component is non--zero.
Right: At an arbitrary intermediate point for the
case $\delta=1$.

\vskip1cm
\centerline{\large \bf Figure 3.1}

\noindent
The behavior of the energy $E_{\rm det}$ (3.24) in the
complex plane for the parametrization $z=\delta e^{i\varphi}$.
The starting point $\varphi=0$ is indicated. From there the
energy continues anti--clockwise.

\vskip1cm
\centerline{\large \bf Figure 5.1}

\noindent
Left: The total energy as a function of the scaling parameter $\lambda$
defined in eqn (5.17)
Right: The parametrical dependence of the energy on the parameter
$R_m$ as defined in eqn (5.18).

\vskip1cm
\centerline{\large \bf Figure 5.2}

\noindent
The profile functions which minimize the energy functional
(5.14) under variation of the scaling parameter
$\lambda$. The quark baryonic density $b\sim q^{\dag}q$ ({\it RHS}
of eqn (5.6)) is artificially scaled such that
the spatial integrals over $\omega/m$ and $b$ coincide.
\vfil\eject

\end{document}